\documentclass[12pt]{aastex}
\usepackage{graphicx}
\begin{document}

\shorttitle{}
\shortauthors{}


\newcommand{\ms}{$M_{\odot}$}
\newcommand{\msb}{$M_{\odot}$~}
\newcommand{\al}{$^{26}$Al}
\newcommand{\fe}{$^{60}$Fe}
\newcommand{\be}{$^{10}$Be}
\newcommand{\ca}{$^{41}$Ca}
\newcommand{\mn}{$^{53}$Mn}
\newcommand{\pd}{$^{107}$Pd}
\newcommand{\tc}{$^{99}$Tc}
\newcommand{\pu}{$^{244}$Pu}
\newcommand{\hf}{$^{182}$Hf}
\newcommand{\ct}{$^{13}$C}
\newcommand{\ctb}{$^{13}$C~}
\newcommand{\li}{$^{7}$Li}
\newcommand{\fl}{$^{19}$F}
\newcommand{\cd}{$^{12}$C}

\title{Deep Mixing in Evolved Stars.\\
II. Interpreting Li Abundances in RGB and AGB Stars}

\author{S. Palmerini}
\affil{Dipartimento di Fisica, Universit\`{a} di Perugia, and INFN, Sezione di Perugia, Via Pascoli, 06123
Perugia, Italy}

\author{S. Cristallo}
\affil{Departamento de F\'{\i}sica Te\'{o}rica y del Cosmos, Universidad de Granada, Campus de Fuente Nueva, 18071 Granada, Spain}
\affil{Osservatorio Astronomico di Teramo, INAF, Via Maggini, 64100 Teramo, Italy}

\author{M. Busso}
\affil{Dipartimento di Fisica, Universit\`{a} di Perugia, and INFN, Sezione di Perugia, Via Pascoli, 06123
Perugia, Italy}

\author{C. Abia}
\affil{Departamento de F\'{\i}sica Te\'{o}rica y del Cosmos, Universidad de Granada,
Campus de Fuente Nueva, 18071 Granada, Spain}

\author{S. Uttenthaler}
\affil{University of Vienna, Department of Astronomy,
T{\"u}rken\-schanz\-stra\ss e 17, A-1180 Vienna, Austria}

\author{L. Gialanella}
\affil{Dipartimento di Scienze Ambientali, Seconda Universit\`{a} di Napoli, Via Vivaldi 43, 81100 Caserta and INFN, Sezione di Napoli, Italy}

\author{E. Maiorca}
\affil{Dipartimento di Fisica, Universit\`{a} di Perugia, and INFN, Sezione di Perugia, Via Pascoli, 06123
Perugia, Italy}

\begin{abstract}
We  reanalyze the problem of Li abundances in red giants of nearly solar metallicity. After an outline of the problems affecting our knowledge of the Li content in low-mass stars ($M \le 3 M_{\odot}$), we discuss deep-mixing models for the RGB stages suitable to account for the
observed trends and for the correlated variations of the carbon isotope ratio; we find that Li destruction in these phases is limited to masses below about 2.3 $M_{\odot}$. Subsequently, we concentrate on the final stages of evolution for both O-rich and C-rich AGB stars. Here, the constraints on extra-mixing phenomena previously derived from heavier nuclei (from C to Al), coupled to recent updates in stellar structure models (including both the input physics and the set of reaction rates used), are suitable to account for the observations of Li abundances below $A$(Li) $\equiv~ \log \epsilon$(Li) $\simeq$ 1.5 (and sometimes more). Also their relations with other nucleosynthesis signatures of AGB phases (like the  abundance of F, the C/O and $^{12}$C/$^{13}$C ratios) can be explained. This requires generally moderate efficiencies ($\dot M \lesssim 0.3 - 0.5  \times 10^{-6} M_{\odot}$/yr) for non-convective mass transport. At such rates, slow extra-mixing does not modify remarkably Li abundances in early-AGB phases; on the other hand, faster mixing encounters a physical limit in destroying Li, set by the mixing velocity. Beyond this limit, Li starts to be produced; therefore its destruction on the AGB is modest. Li is then significantly produced by the third dredge up. We also show that effective circulation episodes, while not destroying Li, would easily bring the $^{12}$C/$^{13}$C ratios to equilibrium, contrary to the evidence in most AGB stars, and would burn F beyond the limits shown by C(N) giants. Hence, we do not confirm the common idea that efficient extra-mixing drastically reduces the Li content of C-stars with respect to K-M giants. This misleading appearance is induced by biases in the data, namely: i) the difficulty of measuring very low Li abundances in O-rich AGB stars, due to the presence of TiO bands; and ii) the fact that many, relatively massive ($M >3 M_{\odot}$) K- and M-type giants may remain Li rich, not evolving to the C-rich stages. Efficient extra-mixing on the AGB is instead typical of very low masses ($M \lesssim$ 1.5$M_{\odot}$). It also characterizes CJ stars, where it produces Li and reduces F and the carbon isotope ratio, as observed in these peculiar objects.

{\noindent {\bf key words}: stars: abundances -- stars: evolution of -- nucleosynthesis -- stars: RGB and AGB -- light elements.}
\end{abstract}
\section{Introduction}
Li abundances in solar metallicity low mass stars (hereafter LMS) are distinctively peculiar
and their spread reveals the inadequacy of standard stellar codes, not including rotation
and assumptions for non-convective forms of mixing. $^7$Li is produced from electron captures
on the short-lived $^7$Be ($t_{1/2}$ = 53 days in laboratory) and, due to its fragility,
is sensitive to any mixing event connecting the envelope  with warm layers ($T \ge 2.5 - 3 \times 10^6$ K), where it easily undergoes p-captures; its production and destruction are also critically
dependent on the H-burning rates controlling the temperature stratification of a star.

A large number of observational and theoretical works tried to interpret the observations of Li
on the Main Sequence (hereafter MS) and on the Red Giant Branch (RGB), for both Population II and Population I stars. By contrast, the number of studies addressing its behavior in the stages of the Asymptotic Giant Branch (AGB), characterized by the presence of thermal instabilities, or {\it pulses} (TPs) from the He shell, is smaller and the observational data are less numerous. Nevertheless, these stages play a crucial role because they show us the integral of all the complex phenomena affecting the Li concentration throughout the evolution of a star. They are, moreover, the site where fresh nucleosynthesis occurs in He-rich layers coupled to convective mixing episodes, the so-called third dredge-up (TDU), increasing the envelope abundance of carbon (up to C/O $>$ 1 for C stars), of fluorine, of technetium and of other neutron-capture isotopes; these independently-produced nuclei can provide tools to understand what is happening to Li. In this paper we plan therefore to reanalyze the abundances of Li in evolved stars, with emphasis on the AGB stages.

Non-convective transport of H-burning products was suggested by many authors
to be required for explaining the Li abundances and other chemical
anomalies in evolved stars \citep[see e.g.][]{was95,sb99,cdn98,b+07,b+10,clan10}. Various physical processes have been explored for
driving such a transport: reviews can be found in \citet{pinson97} and \citet{talon08}.
Special attention was dedicated, in recent years, to thermohaline diffusion
and rotational shear \citep[see e.g.][]{chala10}.


The mixing phenomena induced directly by rotation were shown by
\citet{pala6} to be rather ineffective in evolved stars and the recent analysis by
\citet{chala10} specifies that their role is mainly that of accounting for
star-to-star scatters in Li abundances and of favoring an early contamination
of the envelope with the materials brought up by diffusion mechanisms.
Rotational mixing can be instead important on the MS, where its low
efficiency is compensated by the long available time scales \citep{clan10}.

On the other hand, thermohaline diffusion, as discussed in \citet{egg1,egg2} and
in \citet{cz07}, has the advantage that it descends from an understood
mechanism, namely $^3$He burning into $^4$He and two protons, reducing the molecular weight
and hence inducing mass readjustments. For this reason, it has rapidly become the most
commonly assumed process of transport. However, recently severe doubts have been
advanced also on its effectiveness \citep{den10,dm10}; the inferred velocities of the transport
(and/or the aspect ratios of the salty "fingers" found in hydrodynamical simulations)
were found to be too small by large factors to yield significant changes in
the envelope abundances. This subject must be clearly examined with different hydrodynamical
codes by others before final conclusions can be drawn. Nevertheless we note that a recent
study of the globular cluster M3 confirms that high mixing velocities and large aspect ratios
of the mixing structures, beyond the reach of thermohaline diffusion, are required \citep{angelou}.


In the above uncertain situation, a few recent papers suggested that magnetic fields
might play a role in stellar mixing through the buoyancy of light magnetized bubbles,
either alone \citep{b+07,nor08} or in synergy with thermohaline diffusion \citep{den09,dm10}.
Qualitative arguments in favor of this scenario have been outlined by
\citet{b+10} and by \citet{p+11}, hereafter Paper 1. In any case, the above discussion makes
clear that attributing the abundance anomalies of evolved stars to a specific physical
mechanism might still be a long shot. Hence we prefer to adopt a parametric model
similar to the ones used in \citet{nol03} and in Paper 1, in an attempt to verify
if mixing episodes of whatever nature (characterized by the parameters derived in
Paper 1 from an explanation of CNO isotopic admixtures in evolved stars) can also account
for the behavior of Li.

The present work is organized as follows. In Section 2 we recall the general history
of the Li evolution in LMS  before the AGB phase. We also present a set of relevant
observational data derived from the current literature, which can be used to
constrain the nucleosynthesis calculations. In this way we show that deep-mixing
schemes based on a range of (moderate) circulation rates can easily account for
most of them. In Section 3 we discuss observations on both O-rich and C-rich
TP-AGB stars. In Section 4 we address some issues related to the time
scales for mixing and for $^7$Be decay, which are relevant for Li production and destruction.
Then Sections 5 and 6 afford the interpretation of Li abundances and of their
relations with other observed nuclei in O-rich and C-rich AGB stars, respectively. For this
task we use the stellar models and the extra-mixing scheme already discussed in
Paper 1. Some general conclusions are then drawn in Section 7. We also add, in Appendix I,
a brief outline of our knowledge on the most relevant reaction rates for our analysis.

\section{Li Abundances and the Evolution of Low Mass Stars}

\subsection{The evolution of Li before the TP-AGB Phase}
As mentioned, Li (like other light nuclei, e.g. D, $^{9}$Be, $^{10,11}$B)
is very fragile, so that its abundance in stellar envelopes is easily decreased by any
mixing process, especially convection \citep{boes05}. In canonical stellar models Li is predicted
to be destroyed during the pre-MS phases in stars of low mass, characterized by large convective
envelopes \citep{pinson97,sestito}. In K and M dwarfs the destruction is expected to continue
on the MS, while for higher surface temperatures the convective envelope progressively shrinks
and does not include any more zones hot enough to affect Li. For them, no further change
in Li abundances is predicted before the RGB phase.

Contrary to the above indications from Standard Stellar Models (SSTMs), MS stars of the Galactic disk,
including the Sun, reveal further Li-depleting processes. In the Solar photosphere, for example,
Li is 100 times less abundant than in meteorites \citep{asplund2}. Moreover, in F dwarfs of the Galactic disc a strong decrease of Li abundances is observed \citep[the so-called {\it Li-dip}, see e.g.][]{bta, btb,balacha, boes98}. An interpretation in terms of diffusive mixing interacting with the rapid shrinking of the convective envelope was early advanced by \citet{mic86} and \citet{mc91}.

All this clarified that several problems still exist in solar physics \citep[see][for a review]{sere10}; some have been exacerbated with the recent updates of the solar abundances by \citet{asplund1} and \citet{asplund2}; see also \citet{caffau}. Indeed, the new data require revisions in the same physical inputs of SSTMs  \citep{dp06,cas07}. In recent years \citep[see e.g.][and references therein]{tbl94,bsp04,psc07}, a considerable amount of work has been spent in looking for missing or poorly-treated physics in SSTMs capable of accounting for the observed effects and solve the existing inconsistencies. This includes consideration of rotationally-induced mixing \citep{chala10} and of its interactions with atomic diffusion \citep{vm98}, as well as active magnetic dynamo processes \citep{egg10}, gravitational waves \citep{ct05}, radiative acceleration (i.e. the transfer of extra momentum from the radiation flux to matter) and a treatment of the partial ionization in the stellar interiors \citep{tur98,tc05,ptc07}.

After the turnoff point (TO), low mass stars evolve to the RGB phase, where Li is observed to
vary remarkably, with a general decrease appearing after the H-burning shell has erased the
chemical discontinuity left behind by the first dredge up (FDU). This is the phase where a bump
in the Luminosity Function (LF-bump) appears. Concerning the prediction of the initial Li concentration
at the base of the RGB, we rely on models for the MS and the
Sub-Giant (SG) phases. In particular \citet{chala10} found that, when rotation is not included in the
computations of the previous evolution, for the masses considered here (from about 1.3 to about
3 $M_{\odot}$, representative of the majority of population I RGB stars) the Li abundance
attains values in a small range ($A$(Li) = 1.2 -- 1.5).  Using
a different stellar code \citep[the present version of FRANEC, see e.g.][and references therein]{cri9}, in the calculations performed for Paper 1 we could confirm these findings very well. Assuming slow rotation on the MS, \citet{cm11} found similar values in their models of low mass stars, suitable for the members of the open cluster M67.

When, instead, rotational mixing is included, for $M >$ 1.5 $M_{\odot}$ \citet{chala10} found increasing Li destruction for increasing mass.  One has to notice, however, that the velocities they adopt on the MS are high, between 110 and 300 km/sec. Even in lower masses (their 1.25 $M_{\odot}$ model star) the minimum rotational velocity they consider for the ZAMS is 50 km/sec; on the contrary, \citet{cm11} exclude ZAMS rotational velocities in excess of 20 km/sec in M67, for which they assume a TO mass of 1.3 $M_{\odot}$. These differences in the assumptions are rather common in the literature and make clear how
many uncertainties affect the previous evolution of presently-observed red giants. This is so especially for field stars, where the determination of the mass is made difficult by the crowding of the RGB zones of the HR diagram, where stars of different masses cross the same ($L,~T_{eff}$) regions in different evolutionary phases. This difficulty extends to stars belonging to later stages of evolution. Recently, \citet{bkrl} identified new Li-rich K giants and attributed them to the clump phases, while stars with Li-enhancements measured by \citet{br89} were attributed to the early-AGB stages by \citet{chab00}.

With the nuclear reaction rates discussed in paper 1 and the present physical assumptions of the FRANEC code we actually find that the LF-bump is reached near the RGB tip already for a 2.3 $M_{\odot}$ star. Hence, more massive models would erase the chemical discontinuity created by FDU only
during core-He burning \citep[see also][]{clan10}. Only after this moment deep
mixing phenomena can occur, so that we find that Li destruction in RGB phases is limited to low mass
stars. High masses ($M \gtrsim 2.3 M_{\odot}$), retaining the Li abundance they had at FDU, burn He
at relatively high $T_{eff}$ values; then they move back toward cooler zones in a $A$(Li)-$T_{eff}$ diagram, crossing areas occupied by lower masses on the RGB. The level of confusion is therefore remarkable.

Our post process code for deep mixing calculations \citep[called $MAGIC$, see][]{awr,st09} uses as inputs full stellar evolutionary sequences
calculated with the FRANEC code \citep{cri9,cris11}. These stellar models start from the pre-MS phase, evolve trough core H- and core He burning
(passing trough the RGB stages), and are stopped at the tip of the AGB phase.

\subsection{Deep Mixing on the RGB as the Source of the Spread in Li}
The above problems limit the level of detail with
which one can explain the behavior of Li in large
samples of field RGB (and clump) stars. This is one of  the reasons why we decided to focus our analysis
mainly on AGB phases, where the understanding of Li, and of CNO isotopes affected by deep mixing can be
aided by the presence of fresh nucleosynthesis products that are generated independently (e.g. F, Tc, and
the other $s$-process nuclei, produced in the He- and C-rich layers).

In this Section, dealing with the previous stages, we can only compare the observed
data with very limited modeling. The calculations refer to phases after the LF-bump,
when the natural barrier against mixing created by chemical gradients has been erased.

First of all, we are forced to consider a wide initial spread of Li abundances at FDU, namely that covered
by the values of $A$(Li) found with and without rotation by either our models or those mentioned in the
previous Section. This spread covers roughly 1.8 dex (with $-$0.3 $\lesssim$ $A$(Li) $\lesssim$ 1.5,
excluding extreme conditions for very fast rotation on the MS). This is an unfortunate situation, but
cannot be avoided, at least until better knowledge of MS rotational rates will constrain more firmly the
effects of their induced mixing in dispersing the otherwise rather well-defined Li abundances at FDU. On
the other hand, such a huge interval seems actually to have its counterpart in the observational spread
after the LF-bump, despite the mentioned difficulties in specifying exactly the evolutionary stage for
field RGB stars.

With the above limits in mind, we must verify whether the parametric mixing models already shown in Paper
1 to account for the CNO anomalies can also explain the huge spread in Li abundances and their observed
trend with effective temperature. Figures 1 and 2 show a rather wide sample of Li data from various authors,
referring to evolved stars in the $T_{eff}$ range covering the RGB stages. There was no real attempt to make the
sample complete beyond any doubt; however, it covers the whole relevant area and comes from some well-known
studies on the subject; hence, the data plotted should be typical and there should be no extra bias in
the analysis coming from possibly missing data. Stellar models predict that RGB stars below 3$M_{\odot}$
stay at the cool side of the long-dashed vertical lines, roughly defining the position of the clump (core-He
burning) for different masses. The horizontal line in the Figures labeled "FDU" represents the maximum Li
abundance expected at FDU, when other mixing phenomena (like rotation) do not play a significant role.

\begin{figure}[h!!]
\centerline{{\includegraphics[height=9.8cm, width=9.2cm]{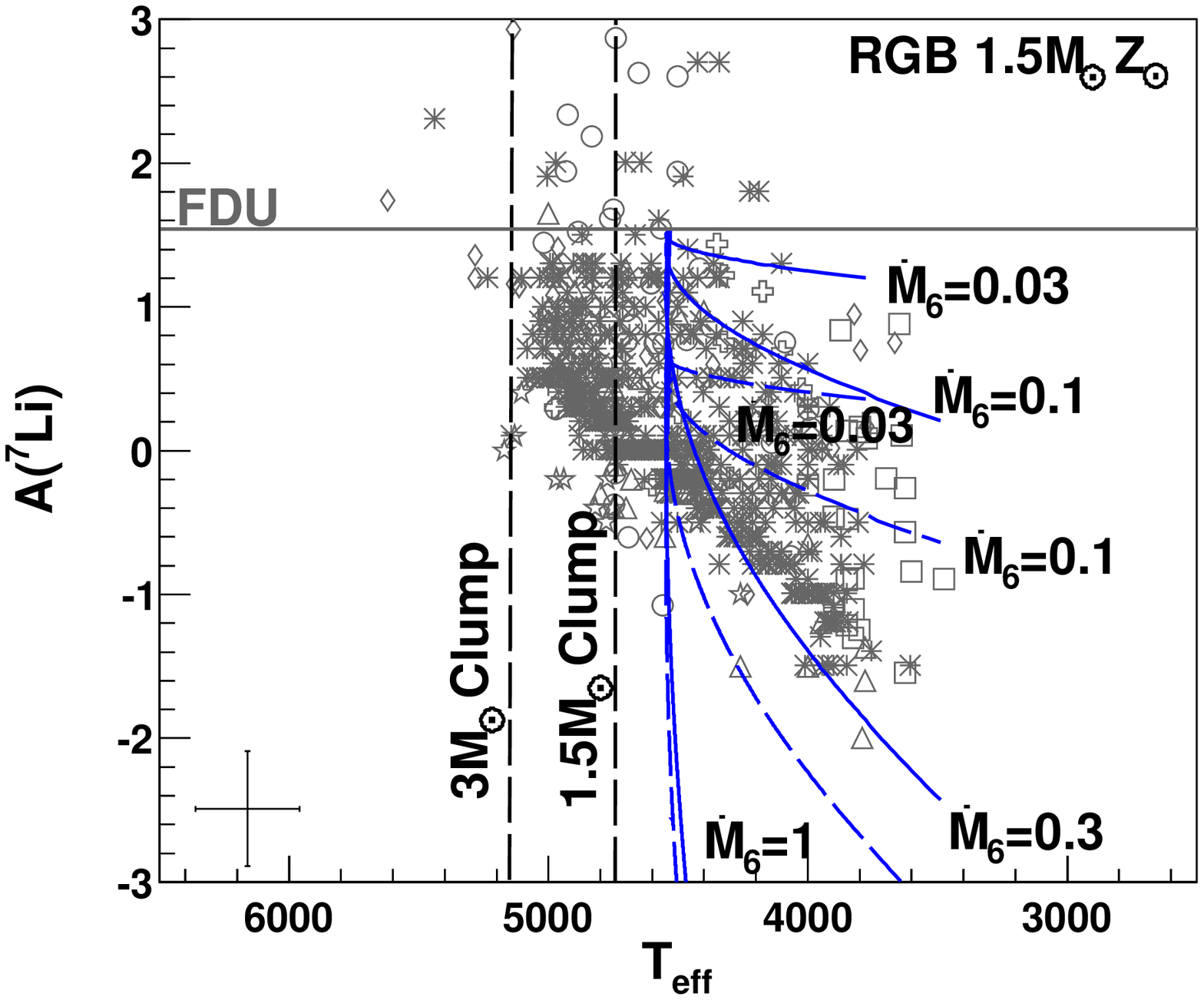}}}
    \caption{A comparison between observations and model sequences for Li abundances in post-Main Sequence Stars.
    Different symbols indicate data taken from different sources. In particular, open crosses refer to data
    from \citet{l77}, open triangles are from \citet{lam80}, asterisks from \citet{br89}, "plus" signs from
    \citet{gil89},  open circles from \citet{mc91}, open squares from \citet{l99}, rhombs from \citet{mal99}, open stars from \citet{cm11}. Model curves refer to extra-mixing calculations for a 1.5 $M_{\odot}$ star of
    solar metallicity.}
              \label{lifig1}%
              \end{figure}

\begin{figure}[ht!!]
\centerline{{\includegraphics[height=9.8cm, width=9.2cm]{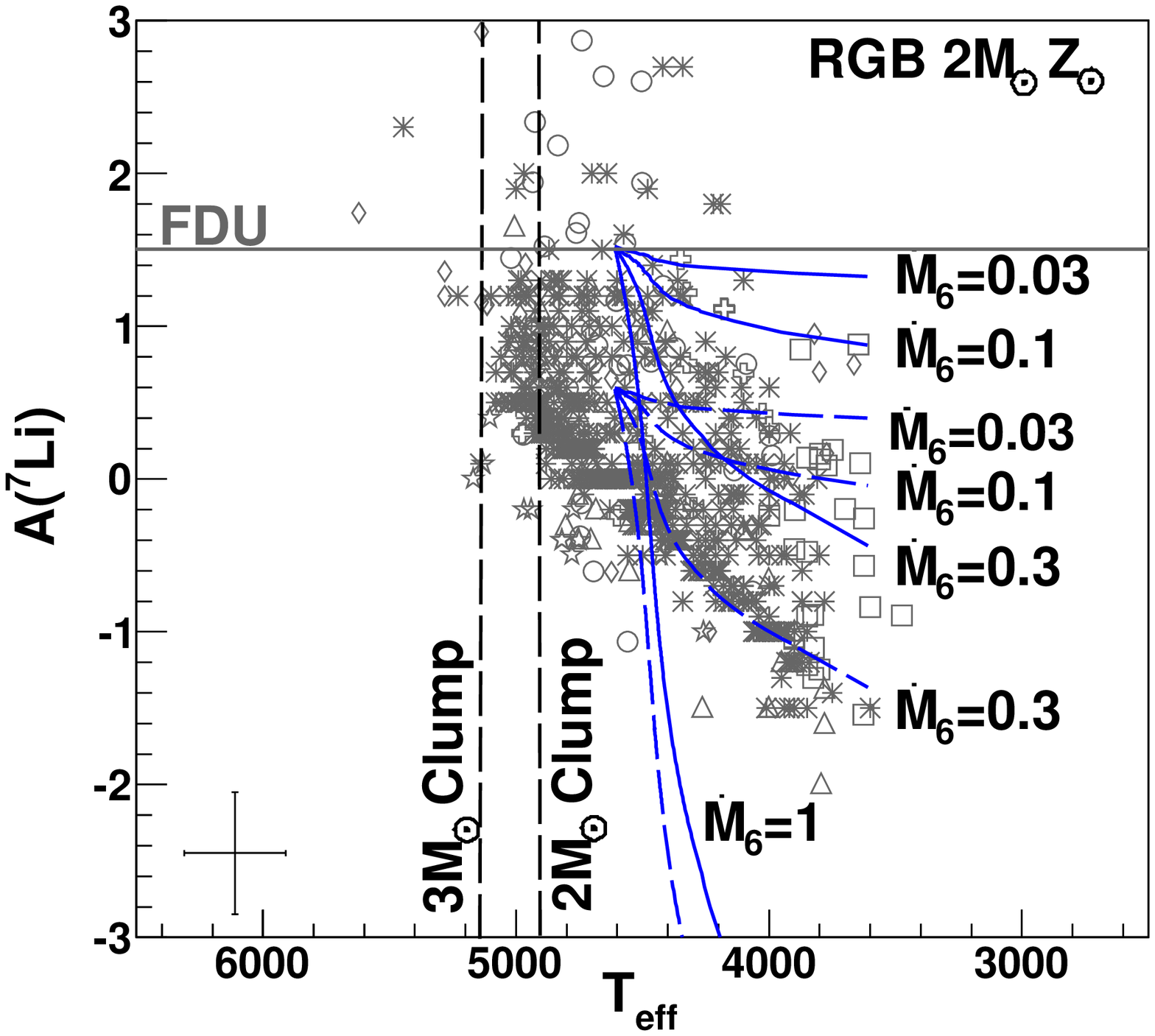}}}
    \caption{Same as Figure 1, but comparing the observations with models for a 2 $M_{\odot}$ star of solar
    metallicity. The same assumptions have been made for the Li abundances at FDU.}
              \label{lifig2}%
    \end{figure}

The two plots contain a series of curves, referring to 1.5 and  2$M_{\odot}$ model stars of solar
metallicity, starting at the temperature for which we find the LF-bump in the two cases. The continuous and
dashed lines  refer to the models discussed in Paper 1, with a parametric deep-mixing process reaching down
to regions where the temperature is defined by $\Delta \equiv \log T_H - \log T_P$ = 0.22. Here $T_H$ is the
temperature at which the maximum energy is generated by H-burning, and $T_P$ is the temperature of the
deepest layer reached by extra-mixing.

\begin{figure}[ht!!]
\centerline{{\includegraphics[height=7.8cm, width=7.5cm]{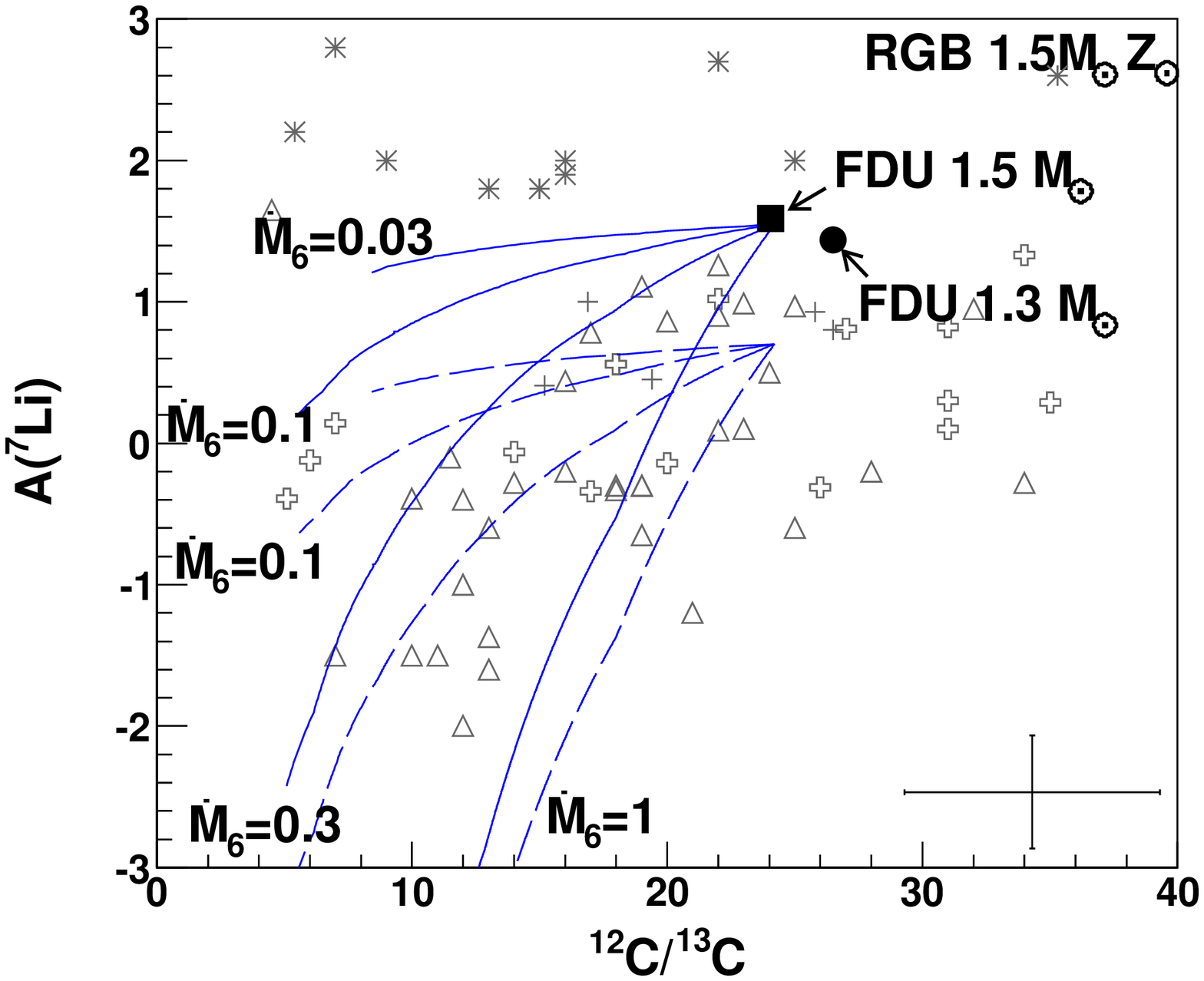}
\includegraphics[height=7.8cm, width=7.5cm]{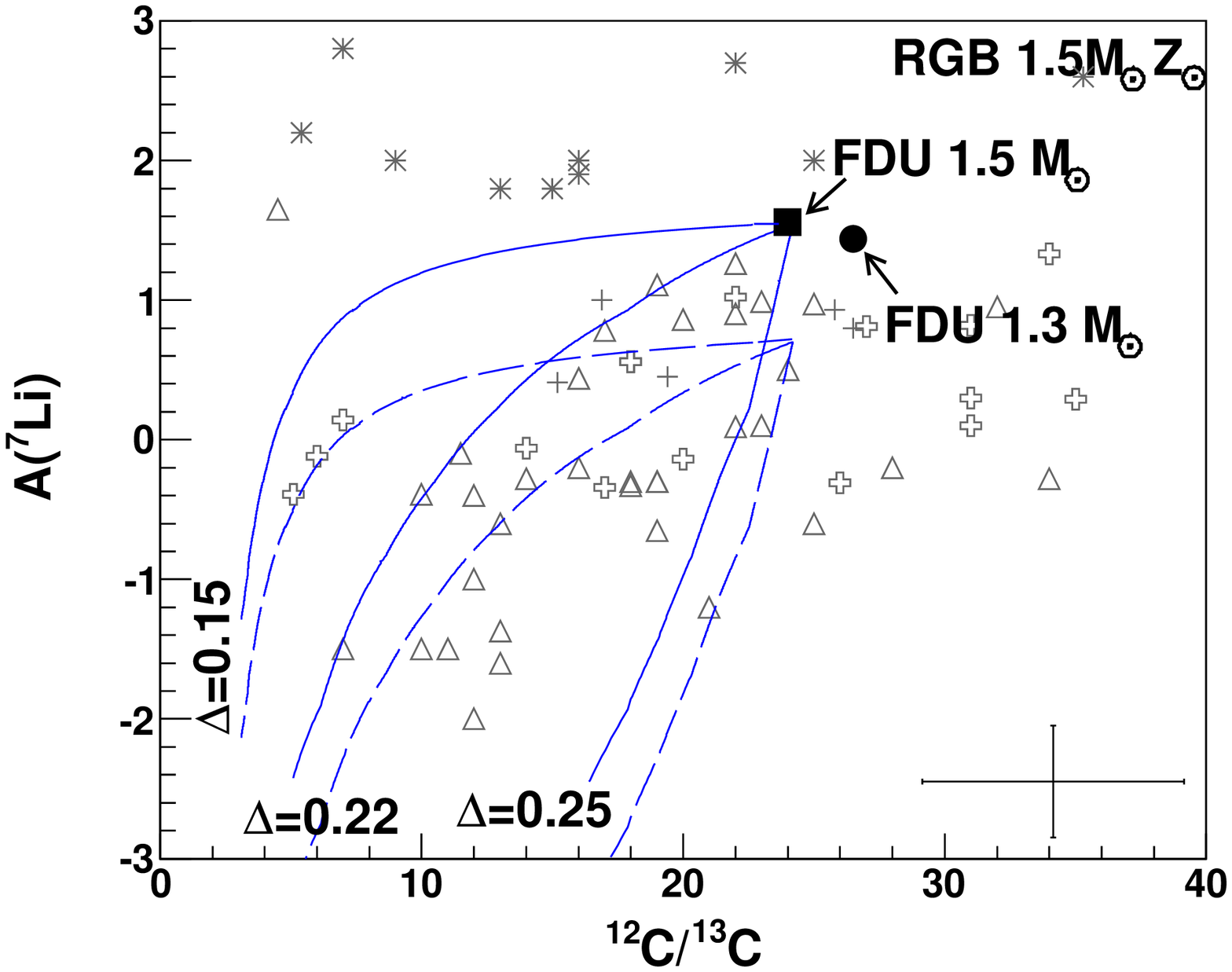}}}
    \caption{The Li abundance as compared to the $^{12}$C/$^{13}$C ratio for those stars, in the sample previously shown in Figures 1 and 2, for which we could find measurements of the carbon isotope ratio. Different symbols again come from different sources and have the same meaning as in Figure 1, apart from
    the fact that the $^{12}$C/$^{13}$C ratios for the stars by \citet{br89} (asterisks) were derived by \citet{chab00}. Model curves are from deep mixing calculations relative to a 1.5 $M_{\odot}$ star of solar metallicity. In the
    left panel we show, for the two Li abundances at FDU discussed in the text, the
    results of computations with $\Delta$ = 0.22 and different circulation rates
    $\dot M_6$, as indicated. The right panel instead keeps $\dot M_6$ fixed at
    the value 0.3 and shows the effects of mixing at different depths, i.e. different
    $\Delta$ values, as indicated.}
              \label{lifig3}%
    \end{figure}

As mentioned above, the starting points of our theoretical curves have to cover a rather wide range.
Continuous lines start from a Li abundance typical of pure FDU, dashed lines from an abundance that would require some rotational mixing on the MS, as inferred from \citet{chala10}.
As in Paper 1, $\dot M_6$ represents the circulation rate adopted, in units of 10$^{-6}$ $M_{\odot}$/yr.
The figures show how deep mixing at the rates already adopted in Paper 1 can
cover the observed spread of Li abundances in RGB stars. Comparing models of different masses in Figures 1 and 2, we can see that in the
1.5$M_{\odot}$ model a drop in Li abundances larger than for the 2$M_{\odot}$ model occurs, for the same choices of the mixing parameters.
This is due to the fact that the time available for deep-mixing during the RGB phase is larger for lower masses.

In the figures, points that stay at the left side of model curves (close
to or inside the clump region) can in this framework be interpreted as representing stars that have terminated the
RGB stage. They have moved toward central He burning, at higher $T_{eff}$ values (and at constant Li) or have already
started climbing the early-AGB track. In our models these phases are, for the major part,
not affected by further mixing, as the shift to higher temperatures corresponds to a retreat of the convective envelope, which
then separates it from the H-burning shell by too large mass intervals (at lest a few $\times$10$^{-2}$ $M_{\odot}$).
Only near the end of the early-AGB stage the envelope mass gets close enough to the H-burning shell
($\Delta M \lesssim 2\times10^{-2}$ $M_{\odot}$), and deep mixing can
become effective again. Apart from these final stages, the model curves after the RGB would therefore be
horizontal lines, moving toward the left and then coming back to the right. They are omitted simply to avoid
excessive crowding in the plots.

\begin{figure}[ht!!]
\centerline{{\includegraphics[height=7.8cm, width=7.5cm]{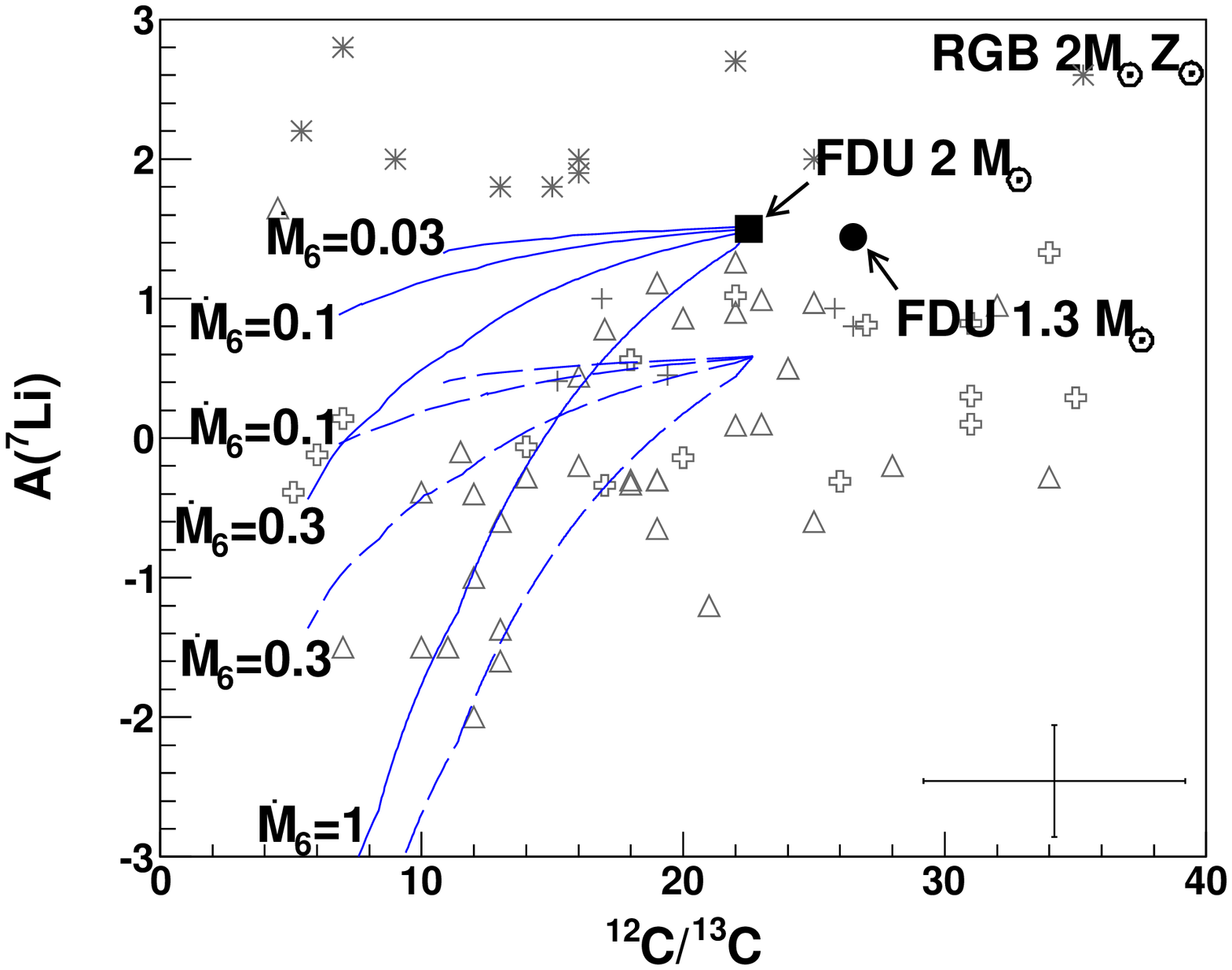}
\includegraphics[height=7.8cm, width=7.5cm]{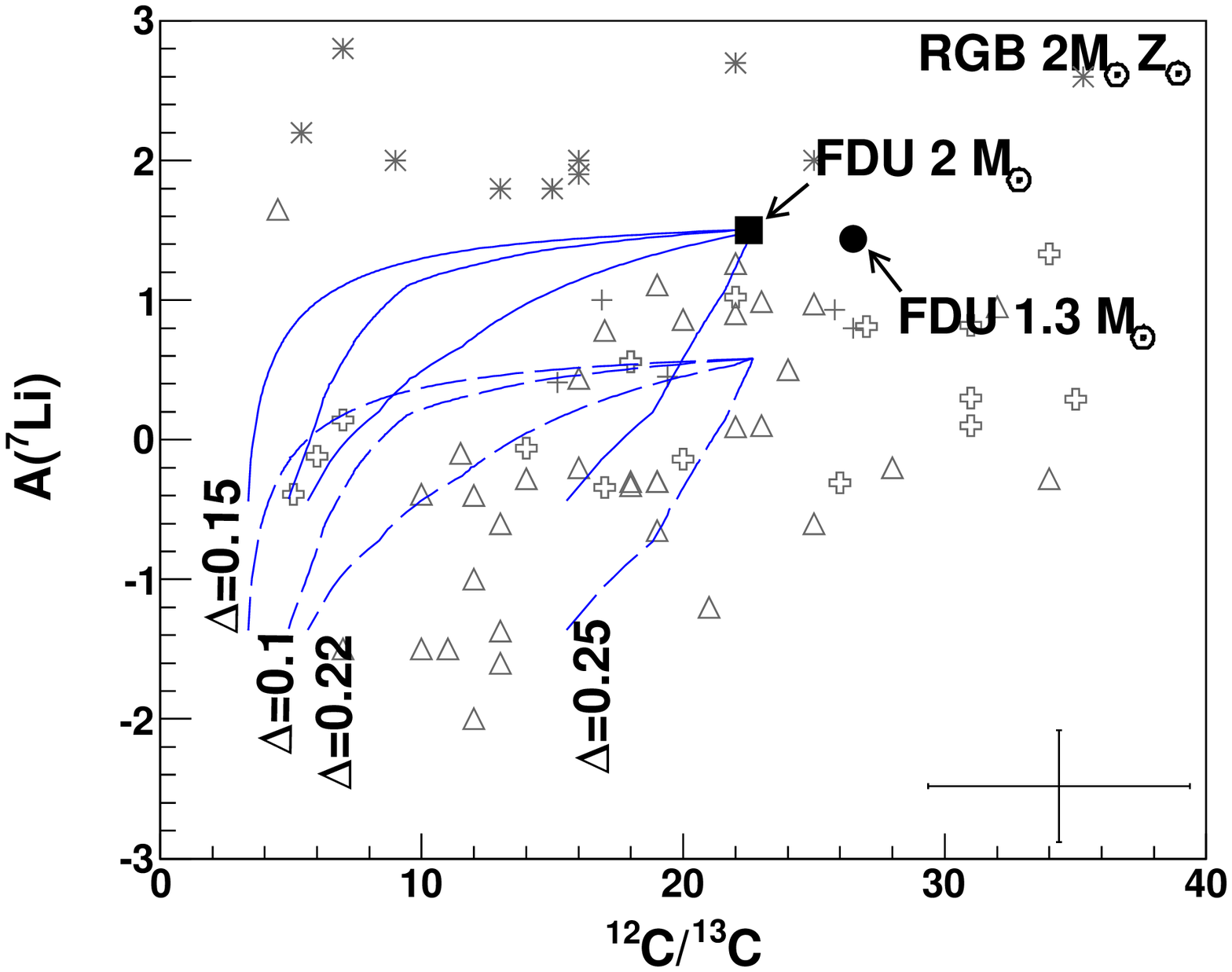}}}
    \caption{Same as Figure 3, but adopting models for a 2 $M_{\odot}$ star of solar metallicity. The meaning of
    symbols and curve tracks is unchanged.}
              \label{lifig4}%
    \end{figure}

The only other chemical constraint to deep mixing on the RGB for which a sufficient number of observations exists is the $^{12}$C/$^{13}$C ratio. Figures 3 and 4 show the relation between this ratio and the Li abundance for a set of stars having observations for both parameters. The curves are from the same models of Figures 1 and 2. It is clear that deep mixing can in general account for these two constraints together.
In Figures 3 and 4 there are several data points with carbon isotopic ratios larger than 25-28. These values are compatible with those expected after the FDU, without requiring any further mixing. The parent stars should either have a mass exceeding 2.3$M_{\odot}$ (so that no extra-mixing is expected during the RGB phase) or be in a phase earlier than the LF-bump.
Notice that, for all data points above $A$(Li) = 1.5 we have no explanation in RGB phases, as we are considering mixing processes whose overturn time is on average longer than $^7$Be decay Hence they destroy Li without reproducing it. (See later, Section 4 and Figures 5 and 6, for a discussion on how things change on the AGB; an analysis of the physical origin of Li-enriched stars will be presented in a forthcoming paper).

\section{Li Abundances in AGB Stars}
\subsection{O-rich AGB Stars}
As low- and intermediate-mass stars ascend the AGB, they become cooler and
cooler, and molecular absorption starts to dominate for the later M spectral
types. The TiO molecule has a particularly strong opacity in the visual and near
infra-red spectra of those stars, significantly hampering the measurement of the
Li abundance in them. Hence, Li abundances for oxygen-rich AGB stars (selected
as large amplitude semi-regular or Mira-like variables) are rare in the
literature, and for many stars only (high) upper limits exist.

Nevertheless, for warmer, early M-type stars measurements of the Li-abundance
have been carried out by \citet{LL82}. Some of the giant stars in their sample
might actually be located on the early AGB. The Li abundances found by these
authors were compared to those in cool carbon stars, the suspected descendants
of M-type giants, in a study by \citet{denn}. These authors found that the Li
abundance in M giants (even though for some of them only upper limits exist), was on
average higher than that in carbon stars. In a simple approach this can be
understood if some Li destruction occurs during the evolution on the AGB. This might
indeed be induced by an extra-mixing mechanism acting on the AGB itself.
The question would however remain if this can be efficient enough in the relatively
short time scale needed by an M-giant to become a C-giant. As we shall see later,
the real situation might be much more complex than this simple views can foresee.


As is the case for carbon stars (see Section 3.2), also in the sample of oxygen-rich stars
some objects are known to show a high Li-abundance, and some form of Li production has
to be envisaged for them. A theoretically and observationally
understood process is the so-called hot bottom burning \citep[HBB,][]{Iben}
in stars of mass $4 M_{\odot} \leq M \leq 8 M_{\odot}$. In this mass range, the
convective envelope penetrates the H-burning shell, thus nuclear burning partly
occurs in convective layers. Under these conditions, carbon
is converted into nitrogen by CN cycling, keeping the C/O ratio below 1, and $^7$Li is produced through the Cameron-Fowler mechanism \citep{CF71}.
The HBB process is well established in SSTMs, though the lower mass limit
for its operation at each metallicity is not well constrained and depends on many input
parameters (convection efficiency, equation of state, opacities, mass loss rate).
Luminous stars that probably owe their large Li abundances to the activation of
HBB have been observed in the
Magellanic Clouds \citep{spl} and in the Galaxy \citep{gh7}, although for the
latter ones the luminosities are less certain. In any case, Li abundances in
HBB stars can be very high, up to $A$(\rm{Li}) $\sim 4.5$.

There are, however, also lower-luminosity oxygen-rich AGB stars with a
considerable Li enhancement, though not as high as in the stars undergoing HBB.
Examples have been identified in the Galactic Bulge \citep{Utt07} and in the
disk \citep{UL10,Utt11}. Because of their lower luminosities, the HBB mechanism
likely does not operate in them, and a different process has to be introduced.
As speculated in \citet{Utt07}, a deep mixing mechanism might be at work that
brings enough $^7$Be from near the H-burning shell to cooler layers, where it
can decay to $^7$Li, and hence increase the surface Li abundance. On the other
hand, standard evolution predicts that surface Li enrichment occurs after a TP, without requiring any
extra-mixing, when the envelope dredges up the H-shell layers. Recent examples of such findings can be
found in \citet{Kar10} and \citet{clan10}, but they reflect a standard situation.
The abundance predicted by models, of the order of $A$(\rm{Li})$\sim 1.5$,
is in good agreement with some observed values. However, these observations of Li abundances enhanced up to high values are rather rare, and stars of similar luminosity and pulsation period (hence similar evolutionary state) can have very different Li abundances. The small statistics of Li-enriched stars implies either that Li is enhanced only during a short phase on the AGB, or that very special conditions must be met by a star to get enriched in Li. In any case, a correlation between the presence of the TDU indicator technetium (Tc) and that
of Li has been observed \citep{Van07,UL10}, supporting the model results that predict some Li enrichment by the operation of TDU on the AGB. The real abundances that can be achieved by the combined effects of destruction and production mechanisms on the AGB remain to be verified in detail (see Sections 4 through 6).

We warn that a comparison of the mean Li abundance between cool M- and C-type AGB stars is difficult because, as already mentioned, the detection threshold is so
much higher in the oxygen-rich stars. Nevertheless, in a sample of M-type AGB
stars in the disk \citep{UL10}, a considerable number of stars is found to
contain Li in the range $0.4 \leq$ $A$(\rm{Li}) $\leq 1.1$, a range were
only few carbon stars are found (see below). This again seems to be in line with what was
deduced from K and early-M giants \citep{denn}, but we shall see later which kind of problems
this idea implies (Sections 5 and 6). In Bulge AGB stars \citep{Utt07}, much fewer stars are found with
detectable Li abundance, which may be interpreted as a mass and metallicity effect, because the old
Bulge AGB stars are expected to be on average of lower mass than the disk stars. If most of these low-mass
stars burn their $^3$He reservoir already on the RGB, for them no fuel will remain to build up noteworthy
Li on the AGB.

\subsection{Carbon-rich AGB Stars}
The existence of AGB carbon stars showing strong Li-line absorptions has been
known for several decades \citep{mckellar,tpw}. These stars are very rare. As stressed
in the previous subsection, this might be ascribed to the fact that the Li-rich status is
experienced during a short-lived phase and/or in stars fulfilling very rare
internal conditions. One should actually distinguish between two groups of
stars, the first named {\it super Li-rich} carbon stars, the second simply
{\it Li-rich} stars. The former objects show Li abundances exceeding $\sim 3.5$
and are indeed very rare \citep[six objects discovered in the Galaxy so far,][]{abia1}, while the latter show abundances below 3.5 and down to about
$1.0$. They are significantly more numerous. Nonetheless, the overwhelming
majority of the Galactic AGB carbon stars show lower Li abundances, covering the wide range $-1.5\leq$ $A$(\rm{Li}) $< 1$, with a majority of $A$(\rm{Li}) values between $-0.3$ and $-1$ \citep[see also][]{denn,abia}. As mentioned, the current explanation of such C-star observations is that they derive directly from those of K and M stars, once shifted to lower average Li abundances during the evolutionary stages in between, thus indicating further dilution occurring in them. We have already advanced some cautions about this interpretation, which will be specified better in the following Sections.
The existence of Li-rich carbon stars has been also found in external galaxies: the Magellanic Clouds \citep{spl,hatz}, the Draco Dwarf Spheroidal galaxy \citep{dom} and M31 \citep{brc}. This shows that Li-rich phases, although rare, occur for different stellar metallicities; there is unfortunately a too low statistics to perform a comparative study in this sense. In this framework one has to notice that \citet{hatz} concluded that the occurrence of Li enrichment among C(N) stars in the LMC is much rarer than in the Galaxy and they do not find any super Li-rich carbon stars in their sample
of 695 stars; however, they also found that Li enrichment in CJ stars
is five times more common than in C(N) giants.

The difficulty with the presence of Li in AGB C-stars is that the great majority of them have luminosities fainter than M$_{bol}\sim -5$. Very low values of the luminosity should be taken with
care, as a large part of the flux is re-radiated in the infrared at very long wavelengths
\citep[up to 40 $\mu$m and beyond, see e.g.][]{guan}. This fact was often
overlooked in past works, affecting the bolometric corrections. Despite this
caution, the data are in general consistent with the fact that the bulk of
Galactic AGB C-stars are low-mass objects ($\la 3\rm{M}_\odot$). At their
luminosities (or masses), burning at the base of the envelope (HBB) is not
expected (at least for metallicities not very much lower than solar). In fact,
many Li-rich stars found in the Galaxy and in the Magellanic Clouds \citep{gh7,sl89,sl90,psl93,spl}
are O-rich (as expected by the action of HBB) with bolometric magnitudes much brighter
than $-5.0$ and masses in the range $4 \leq \rm{M/M}_\odot \leq 8$ (see above). Thus, a
different mixing/burning mechanism is needed to explain the Li-rich and super Li-rich C-stars, especially of CJ type \citep[as well as to explain Li-enrichment in those O-rich objects for which the luminosity indicates a low mass, see][]{Utt07}. As mentioned, we shall dedicate another work to this problem.

In any case, the situation has been so far rather confusing, because even the Li-normal C-stars are difficult to explain, on the basis of current
models. As mentioned, normal evolutionary calculations of low mass AGBs predict a smooth increase of the
photospheric Li content because the TDU episodes carry to the envelope materials belonging not only to the He-rich intershell zone, but also to the layers above the H shell, where $^{7}$Be is produced. Typically, expectations are of $A$(\rm{Li}) $\sim 1 - 1.5$ after a few TDUs. The production of Li is
roughly proportional to the concentration of $^3$He left in the envelope after
the RGB phase. Except if one assumes that most of the $^3$He is burnt in
previous evolutionary phases, it would appear that mechanisms destroying the
freshly-added Li were required (as said, the majority of C-stars
have Li abundances below $-$0.30). From this discussion, one might apparently
conclude that the so-called (moderately) Li-rich AGB C-stars might be just those stars
where no extra-mixing mechanism destroying Li operates. However, most of
these objects are $^{13}$C-rich, showing a low carbon isotopic ratio
($<25$, Abia \& Isern 1997). Carbon isotopic ratios lower than about 25 in AGB
C-stars of solar metallicity can only be explained assuming some kind of
extra-mixing acting in the same AGB phases \citep{b+10}.

In summary: a) if no $^3$He is left for the AGB phase, one might think that in Li-rich C
stars there is no process affecting Li, leaving it at the level left by the FDU.
However, this is in contradiction with the low $^{12}$C/$^{13}$C ratios, which require extra-mixing.
As mass circulation at rates faster than Be decay produces Li (still destroying $^{12}$C), Li-rich C-stars may possibly experience relatively fast episodes of mass transport. If this evolutionary trend is possible, J-type carbon stars with very low carbon isotope ratios and abundant Li may correspond to extreme examples of this situation. One needs however an explanation for their being s-process poor, contrary to expectations. b) In the case in which some $^{3}$He remains after the RGB phase, then normal (Li-poor) C(N) stars should either have efficient extra-mixing mechanisms destroying Li and compensating for any Li production at TDU, or they may simply descend from Li-poor M giants (where Li has been previously destroyed during the RGB evolution); these M-type progenitors would not be observed for the selection effects introduced by the TiO bands. We shall soon see that this last is actually the only possible scenario.

On the other hand, the recent re-analysis of F abundances in C-stars \citep{abia2} has to be taken
into account, adding more puzzles to this situation (see Section 6 and its Figure 12). C-stars
with normal Li show a range in F enhancements. These stars also have $s-$element enhancements
correlated with F, in agreement with model predictions. Nonetheless, the Li-rich C-stars show no
F enhancement, or even some F depletion. Most of these stars are of J-type not showing s-elements
\citep{abia11}. If this anti-correlation between Li and F is confirmed, then whatever is the real
evolutionary path of case {\it a)} above, it has to destroy $^{12}$C or produce $^{13}$C (explaining the very low carbon
isotope ratios) in stars where F is not synthesized or is destroyed.

\section{On the interplay of nucleosynthesis and mixing for Li on the AGB.}

We adopt here the observations discussed in the previous Section as constraints
to our nucleosynthesis and mixing models  for AGB stars. As already anticipated,
we want in particular to derive clues on the evolution of the Li abundances
using also the independent information provided by nuclei synthesized in the He-
and C-rich layers, like F, Tc, $^{12}$C (through the C/O and $^{12}$C/$^{13}$C ratios)
etc. Conventionally, we assume that extra-mixing can restart when the radiative layers
above the H-burning shell become smaller than 0.02 $M_{\odot}$ (i.e. equal to what
existed at the LF-bump on the RGB). This implies that
only the final part ($\sim$ 1 Myr) of the early-AGB phase can be considered as viable for
deep-mixing activation. Although even before that moment the radiative region below the envelope
is homogeneous in composition, covering its huge extension by transport phenomena would require unrealistic assumptions, e.g. very large mean free paths for the mixed material, or very large diffusion coefficients, depending on the kind of treatment.

\begin{figure}[h!!]
\centerline{{\includegraphics[height=7.8cm, width=7.5cm]{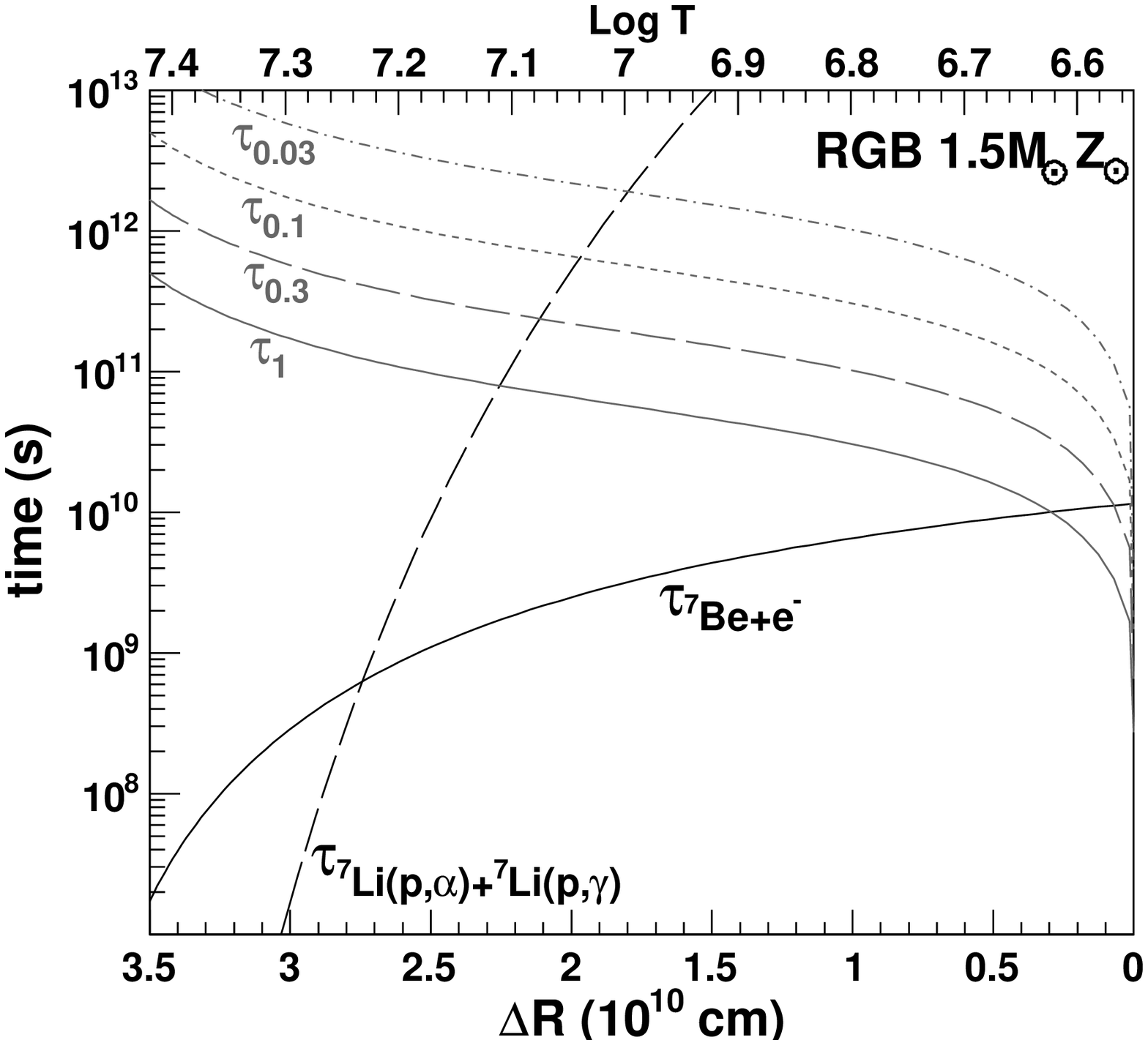}}
{\includegraphics[height=7.8cm, width=7.5cm]{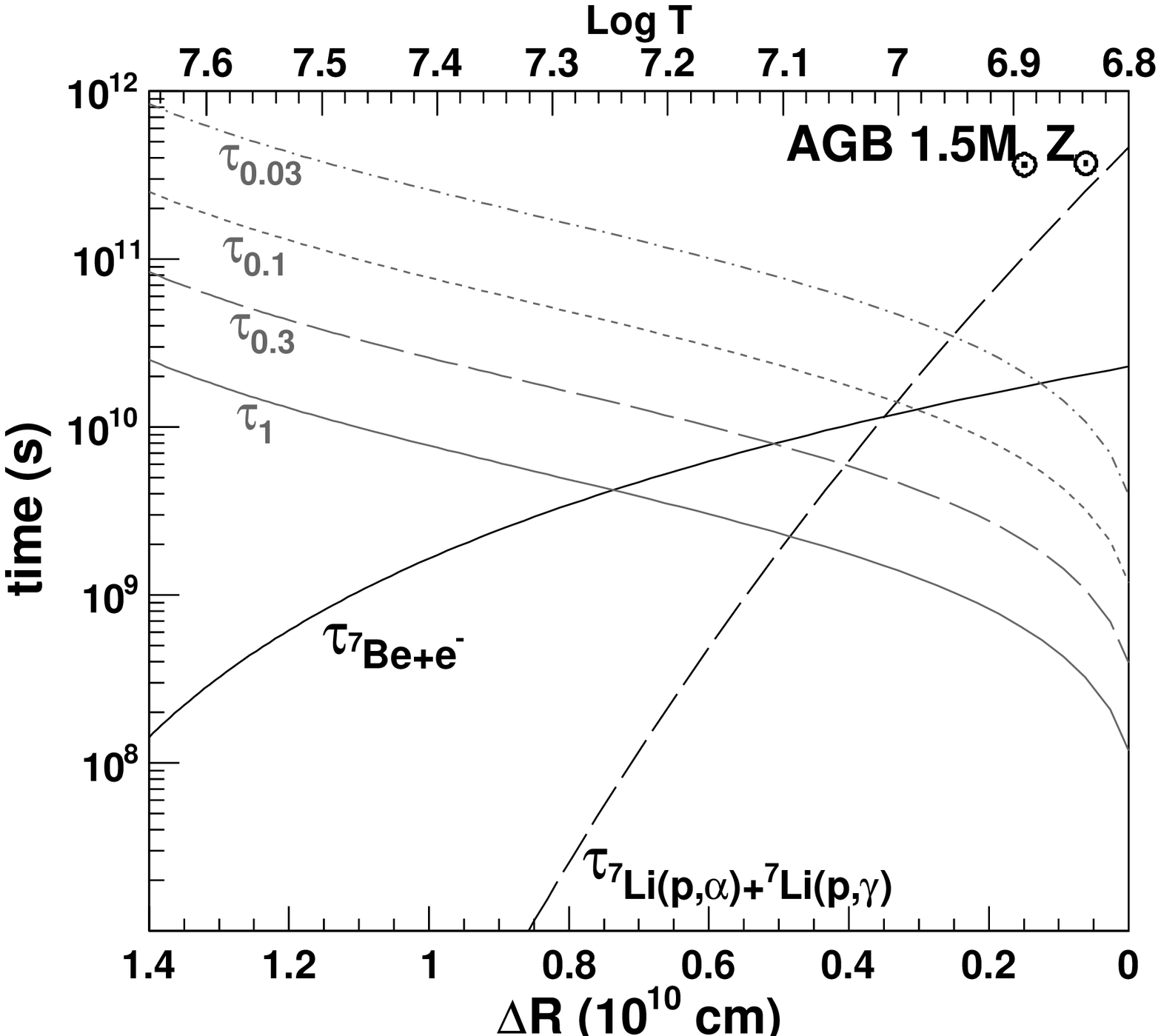}}}
    \caption{A comparison between the time scale for Li
    production from Be decay, that for Li destruction by p-captures
    and the overturn time of mixing for different values of the circulation rate $\dot M_6$. The plots refer to the radiative layer between the H-burning shell and the base of the convective envelope in our 1.5 $M_{\odot}$ model star. On the lower abscissa, $\Delta$R indicates the distance from the base of the stellar envelope; the upper abscissa shows the corresponding temperature. The time scales for mixing are labeled by the letter '$\tau$' with a subscript that indicates the value of the assumed circulation rate $\dot M_6$. Left panel: the RGB case. Right panel: the AGB.}
              \label{lifig5}%
\end{figure}

\begin{figure}[h!!]
\centerline{{\includegraphics[height=7.8cm, width=7.5cm]{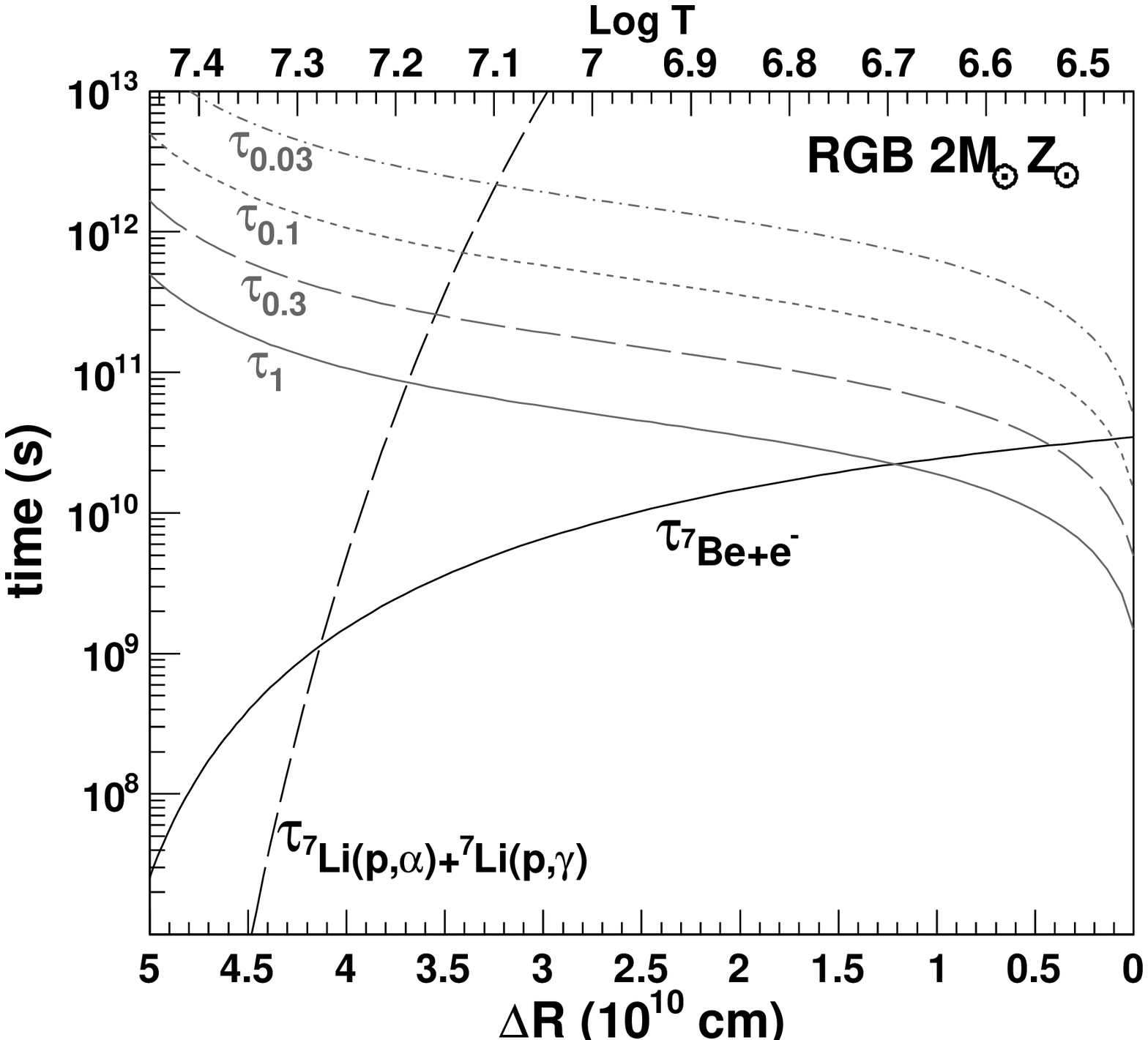}}
{\includegraphics[height=7.8cm, width=7.5cm]{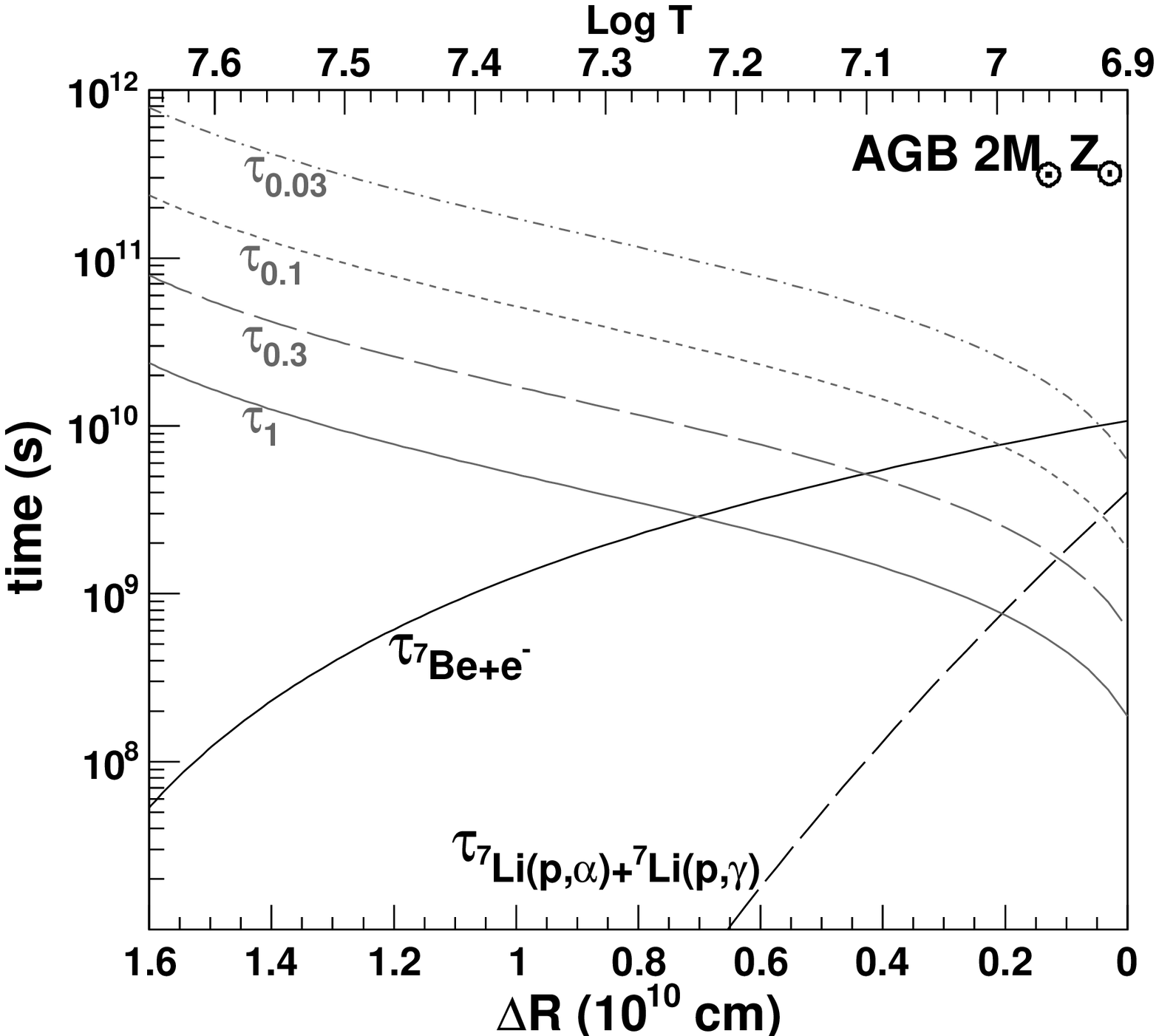}}}
    \caption{Same as Figure 5, but for our 2 $M_{\odot}$ model star}
              \label{lifig6}%
\end{figure}

The observational evidence we have discussed previously for Li is complex, but two main characteristics exist, which were noticed many years ago by \citet{denn}: i) observations in C(N) stars seem to imply, on average, lower Li concentrations than observed in M giants (so that it has been so far a common assumption that deep mixing must destroy Li in between these evolutionary stages); and ii) Li abundances appear to be rather unrelated to the $^{12}$C/$^{13}$C ratios (except for CJ stars). These two results are puzzling, as they appear mutually contradictory (destruction of Li should correspond to a decrease of the carbon isotope ratio).

As is often the case, the contradiction is actually an indication that various non-obvious phenomena are at play. These include the large Li-abundance spread inherited by the RGB and the observational difficulties for O-rich AGB stars mentioned in Section 3.1. They also contain the combined effects of the short half-life of $^7$Be and of the short time (less than 4 Myr) available for extra-mixing on the AGB, including the TP-AGB phase (and in this last case the transport occurs actually in a series of short time intervals in the interpulses, separated by the general reset of abundances operated by TDU).

Let's try to explain this last point in some detail. As the lapse of time is rather short, obtaining large effects on abundances would require circulation rates larger than on the RGB ($\dot M_6 \gtrsim$ 0.3). In paper 1 we showed that this is adequate for explaining CNO isotopes and $^{26}$Al in presolar grains (and actually circulation rates in excess of $\dot M_6 \simeq$ 1 are necessary for them). In principle one could guess that, even for Li, establishing an evolutionary link between relatively Li-rich K and M giants and relatively Li-poor C stars would require a similarly efficient mixing. But this is difficult to obtain in practice. Indeed, mixing effectively in a short time interval also means a relatively fast velocity, to the point that the overturn time scale might become smaller than $^7$Be lifetime. In a simple scheme, one can assume for the velocity of the circulation:
$$
v = {{\dot M}\over{4 \pi r^2 \rho f}}  \eqno(1)
$$
where $f$ is the fractional area occupied by the mixing flow (as in Paper 1, for this example one can assume
$f_{up}$ = $f_{down}$ = $f$ = 0.5). As there is very little mass in the radiative  region, the density roughly follows a $\rho \sim r^{-3}$ law, so that the velocity increases radially. For efficient transport, this implies that above a certain layer mixing becomes faster than $^7$Be decay (see Figures 5 and 6, referring to the 1.5 and 2 $M_{\odot}$ models). In regions external to that point, Li will be no longer destroyed, but produced. As the left panels of Figures 5 and 6 show, on the RGB this is not really a problem, at the typical mixing rates there required (see Section 2). However, for the AGB (right panels) the problem exists (and in the 2 $M_{\odot}$ model is very severe). Then, on one side slow mixing will not change the Li abundance much; on the other, a faster circulation is at risk of reverting destruction into production.

In general, Li destruction on the AGB cannot therefore be too large, but everything will depend on
how much Be has remained in the rising flow. This last parameter is a function of the temperature in
the deepest layers attained by mixing: a shallow circulation ($\Delta$ values in excess of 0.15) mainly
samples the peak of Be production, hence preserves a lot of Be at the moment when the time scale for
mixing becomes shorter than that for Be destruction. In such cases, increasing the mixing rate
$\dot M$ will not be useful to destroy Li anymore; actually it will produce it. A more effective
destruction is obtained by very deep mixing ($\Delta \sim 0.1$): in this case the flow spends more
time consuming Be and therefore reaches the critical point with lower Be abundances; anyhow,
there is always a limit beyond which destruction is converted into production.

\section{Modeling Li abundances in O-rich AGB stars}

With the above limits in mind, let's analyze the situation for O-rich AGB stars.

\begin{table}[t!!]
\centerline{Table 1}
\centerline{Initial conditions adopted on the early-AGB in Figure 7}
\centerline{(1.5 M$_{\odot}$ models of solar metallicity)}
\centerline{
\begin{tabular}{l|c|c|c|c}
\hline
Line & $\Delta_{AGB}$ & $\dot M_{6,AGB}$ & \multicolumn{2}{c}{
$^{12}$C/$^{13}$C after the RGB} \\
\hline
 & & & a & b and c \\
\hline
 continuous & 0.22 & 1 & 8.31 & 8.28 \\
 dashed & 0.22 & 0.3 & 20.89 & 20.43 \\
 dotted  & 0.15 & 0.3 & 38.26 & 36.86 \\
 dash-dotted & 0.1 & 1 & 50.01 & 49.64 \\
\hline
\hline
Case & X($^3$He) & X($^7$Li) & $\Delta$ ; $\dot M_{6,RGB}$ & FDU Estimate \\
\hline
a & 2.79$\cdot 10^{-4}$ & 7.16$\cdot 10^{-11}$ & 0.22 ; 0.03 & FRANEC \\
b & 6.18$\cdot 10^{-5}$ & 7.23$\cdot 10^{-12}$  & 0.22 ; 0.1 & FRANEC \\
c & 6.18$\cdot 10^{-5}$ & 1.03$\cdot 10^{-12}$ & 0.22 ; 0.1 & \citet{chala10}\\
\hline
\end{tabular}
}
\label{ini1p5}
\end{table}

\begin{figure}[t!!]
\centerline{{\includegraphics[height=8.5cm, width=7.8cm]{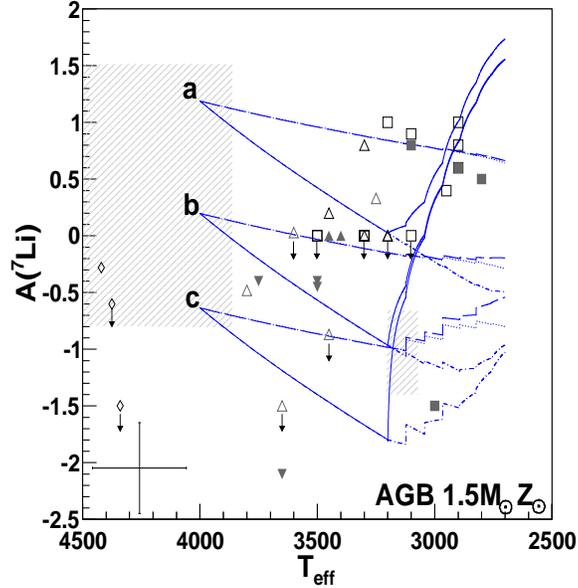}}}
    \caption{The Li abundances observed in O-rich stars on the AGB.
    The stars shown cover the evolutionary phases of the early-AGB (from late K to late M types)
    and of the TP-AGB (MS, S, SC). The data points are taken from the current literature.
    In particular, the shaded area on the left covers the region of the several post-RGB stars
    present in Figures 1 and 2. Then individual observations are indicated as: empty black rhombs for K giants \citep[from][]{lam80,RL05}, empty grey triangles and empty black triangles for M giants and MS stars \citep[from][]{B76,LL82,Van07}, filled grey triangles for S stars and reversed filled grey triangles
    for SC stars \citep[from][]{KW90,Van07}. Some data are taken from \citet{UL10}, and are shown as black empty squares (S
    stars without Tc) or as filled grey squares (S stars with Tc). The shaded area on the right
    side indicates the region where measurements for cool stars quoted by \citet{chala10} lay.
    Model curves are characterized by the parameters described in the text and in Table 1.}
              \label{lifig7}%
    \end{figure}

Our observational constraints are presented in Figure \ref{lifig7}. At the highest temperatures they cover an area (shaded in the figure) where we have several stars already present in Figures 1 and 2, for which we discussed in Section 2 the evidence that they are in post-RGB phases (they are mainly of K-type). We then have M-type giants that are most probably on the early-AGB, plus MS, S and SC giants belonging to the TP stages. The data points are distributed over a wide interval in $T_{eff}$; this indicates that the parent stars have spent a different time on the AGB, as the evolution here proceeds from warm to cool regions of the plot. A group of data for very cool stars is taken from \citet{chala10}; for them, observational details have not been published so far, so that they are again simply represented by a shaded area. They show very small Li abundances and low temperatures. We have to remember, as discussed in Section 3.1, that it is very difficult to measure Li concentrations below about $A$(Li) = 0 for cool O-rich stars; we suspect therefore that most of these measurements are actually upper limits. We notice, in any case, that the cool and Li-poor stars from this sample would roughly correspond to those (many) Li-poor M stars whose existence was noted also by \citet{Van07}, although these authors considered individual abundances as not measurable. These cautions, and the discussion of Section 3.1, make clear that the observational material available for Figure \ref{lifig7} certainly presents selection effects, to which one should pay due attention.

The curves of Figure \ref{lifig7} refer to our AGB calculations, which are based on the 1.5 $M_{\odot}$ FRANEC model with solar metallicity. This model would become C-rich at the last pulse in the absence of deep mixing; but even a minimal occurrence of this last phenomenon prevents it from forming a C-star. In fact, in Paper 1 this model was considered as a prototype for many MS-S giants of very low mass at solar metallicity, never becoming C(N) stars. It confirmed independent evidence already accumulated in recent years on the fact that the mass interval covered by O-rich AGB stars must be larger, and must include lower masses than possible for C stars \citep[see e.g.][]{guan,bu07,Utt07,gb08}. (Obviously, we do not know the mass of the stars plotted in Figure 7. In this Section we compare them with an evolutionary trend that would not lead to a C star. In the following Section they will be compared with model sequences of a higher mass, which do become C-rich).

Since the sensitivity of Li abundances to the stellar conditions is such that it is not easy to illustrate the many ramifications in a simple way, we computed several extra-mixing calculations for the AGB, starting from 3 different sets of  initial  abudances, in which Li, C isotopes and $^{3}$He are distributed over the whole range of observed situations. The specific  $a, b$ and $c$ cases were selected from our subset of extra-mixing calculations on the RGB phase  presented in section 2. For the sake of clarity, we summarize the parameters characterizing the various runs in Table 1.
In our calculation mixing interacts with convective TDU when the TPs start. The different curve tracks shown in Figure \ref{lifig7} for each initial condition ($a, b, c$), refer to different choices for $\dot M_6$ and $\Delta$ on the AGB. As one can see, even this limited set of parameters gives rise to widely different predictions.


\begin{figure}[h!!]
\centerline{{\includegraphics[height=0.6\textheight]{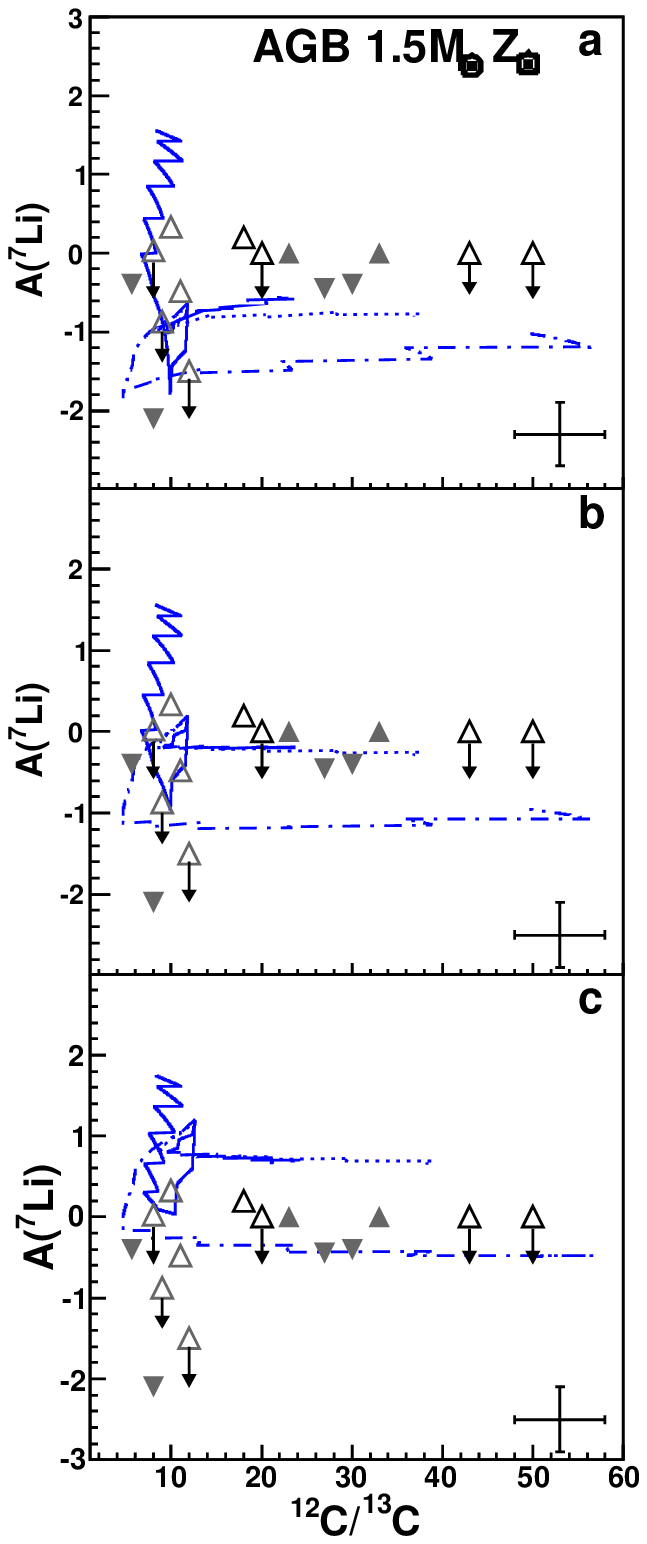}}}
    \caption{The relations between Li abundances and the $^{12}$C/$^{13}$C ratio for O-rich AGB stars.
    The meaning of symbols and the cases shown are the same as in Figure 7. Observations of carbon isotopic ratio are from \citet{JSL92}. In the three different panels the complex paths of the curves start at the $a), b)$ and $c)$ positions, respectively, move backwards (because of $^{12}$C destruction from extra-mixing on
    the early-AGB) and then reverse their trend when TDU starts to
    enrich $^{12}$C in the envelope. See text for further comments.}
\label{lifig8}%
    \end{figure}

As the figure shows, each curve has a first smooth, descending track, corresponding to extra-mixing alone, acting in the early-AGB stage. It is followed by a more complex and step-wise trend induced by the interaction of non-convective mixing (in the interpulse stages) and TDU (after the pulses) along the TP-AGB. In the first phase, Li can be destroyed by deep mixing. Moderate circulation rates ($\dot M_6$ = 0.3, dashed and dotted lines) correspond to moderate Li consumption (up to about 0.4 dex). More extreme cases, with $\dot M_6$ =1, however (continuous and dash-dotted lines, especially for cases $b$ and $c$ seem to allow for a direct evolutionary link between relatively Li-rich K-type stars and Li-poor, more evolved ones, with a trend reminiscent of the suggestions by \citet{denn}. During the early-AGB phases the Li re-production mentioned in Section 4, typical of efficient mixing cases, is rather small and not evident in the plot (although it exists). Subsequently, the remaining $^3$He rapidly induces Li production at TDU. In the case of a shallow extra-mixing (continuous line), the circulation in the interpulse stages contributes to this Li production, thanks to the time-scale effects illustrated in Figures 5 and 6, and high Li abundances are built up. This evolutionary trend therefore ultimately leads to high Li abundances for cool AGB stars, not to low ones, despite the efficient non-convective transport. Anyhow, this case explains a number of observations. For deeper circulation phenomena ($\Delta = 0.1$, dash-dotted line) the Li reproduction at TDU remains moderate and Li-poor evolved stars can be produced, but with some unwanted consequences.

This is shown in Figure 8. There, the same cases of Figure 7 are plotted, but using as abscissa the trend of the $^{12}$C/$^{13}$C ratio. The curves start at the $a, b, c$ points and then move to the left, as extra-mixing on the early AGB  reduces the $^{12}$C/$^{13}$C ratio. In cases with efficient circulation very low values of this ratio can be reached, before TDU increases $^{12}$C and the curves move to the right side of the plot. In the first stage, the cases so far discussed (with $\dot M_{6}$ = 1 for the $b$ and $c$ choices of initial abundances) always lead to a remarkable reduction of the carbon isotope ratio. The continuous tracks of both cases $b$ and $c$ ($\Delta$ = 0.22) spend more than 50\% of their total evolutionary time (and actually most of the TP-AGB phase) with $^{12}$C/$^{13}$C ratios below 10. The deepest case (dash-dotted, $\Delta$ = 0.1) spends in this condition only 25\% of its time, but reaches down to the equilibrium value before the onset of TPs. If these models were typical, hence several M, MS and S stars in Population I should show very low $^{12}$C/$^{13}$C ratios. Some of these stars do exist (as the plot shows) but they are not very common. Notice that subsequently the dash-dotted curve (most efficient and deepest mixing) of case $a$ compares well with several data points, (remember that the values $A$(Li) $\lesssim$ 0 are generally upper limits). The same is in any case true for the more moderate mixing of case $b$ with $\dot M_6$ = 0.3 (dashed line), which does not imply large numbers of $^{13}$C-rich stars.

Summing up, AGB stars of very low mass, never becoming C-rich, can experience significant Li destruction
on the early-AGB; but this implies either producing subsequently Li up to high abundances (continuous curves), or passing through
rather long phases as M, MS, S giants with low $^{12}$C/$^{13}$C ratios, down to CNO equilibrium (dash-dotted curves).
Very efficient mixing seems therefore unlikely, although at this point we do not have yet very strong arguments against this possibility.

\begin{figure}[h!!]
\centerline{{\includegraphics[height=0.6\textheight]{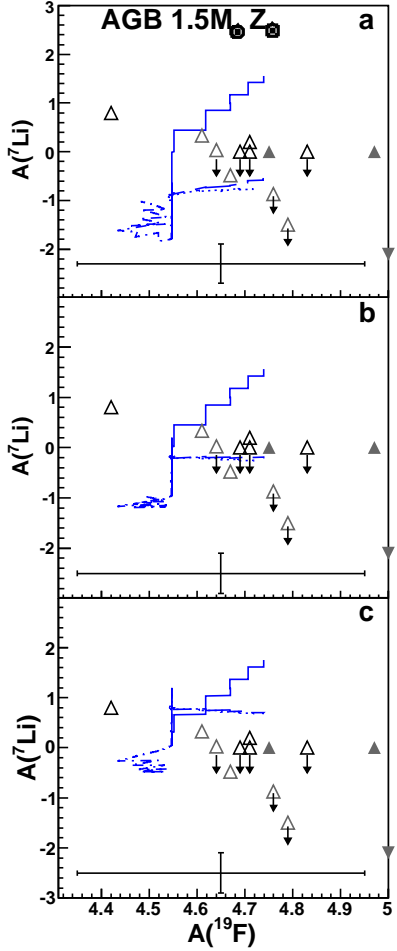}}}
    \caption{The relative trends of Li and F abundances for O-rich AGB stars. For the data symbols and
    the model curves we adopt the same assumptions already made in Figure 7 and Table 1. Observations of F are from \citet{JSL92} and \citet{Utt08}}
              \label{lifig9}%
    \end{figure}

We want now to compare the trend for Li abundances in O-rich AGB stars with independent constraints provided by nuclei produced in the He-rich layers. Not much can be said for Tc, due to the very low statistics. However, it is at least important that Tc-rich S stars (grey squares) do stay in Figure 7 in the region where TPs are active (they start at $T_{eff} \sim $  3200 K).

More relevant information can be obtained through $^{19}$F, a typical stable nucleus synthesized in TPs. We predict the F abundance in the envelope by means of the same models already discussed in \citet{abia2}. This nucleus is  dredged-up to the envelope in adequate concentrations (it is produced mainly by the reaction $^{15}$N($\alpha$,p)$^{19}$F) in stars where the formation of the so-called $^{13}$C-pocket is possible. This is the well-known reservoir of $^{13}$C below the layers achieved by TDU, which also acts as the main neutron source for $s$-processing. For all these model issues we refer to the original works whose treatment we followed \citep[see][for details]{cri9,abia2}.

The comparison of F and Li abundances is presented in Figure \ref{lifig9}. The continuous and dash-dotted tracks ($\dot M_6$ = 1) do not cover useful areas of
the plot. Actually, the deepest and most efficient mixing cases (dash-dotted lines) that couldn't be completely excluded from Li and carbon-isotope data alone
(but raised doubts and warnings) are found to extend to regions so hot that $^{19}$F is partially destroyed (down to $A$(F) = 4.42), leading to predictions
that seem to be in conflict with the data, despite the large error bars. Here we see that the information from He-burning products, despite the low statistics and
huge uncertainties, is helpful. On the basis of the (limited) evidence from F data and its converging indications with previous hints derived from Li and CNO nuclei,
we think we can tentatively exclude, for AGB stars around 1.5$M_{\odot}$, the possibility of very efficient mixing. A comparison with Paper 1 (where such efficient
mass transport was shown to be important in very low mass AGB stars of $M = 1.1 - 1.4 M_{\odot}$, for explaining oxygen isotopes in presolar grains) confirms the common
finding that the effectiveness of deep mixing increases for decreasing stellar mass. Hence, the stars showing F should be more typical of cases in which non-convective
transport on the AGB occurs at moderate rates, as shown by the dashed and dotted curves of cases $b$ and $c$. We have actually verified that a spread in $\dot M$ values
between 0.3 and 0.45 would account for essentially all the data points in Figure \ref{lifig9} within the uncertainties.

If the constraints from C and F can be trusted, than coming back to Figure 7 we should conclude that the evolutionary paths implying remarkable Li destruction, which seemed to be possible for that figure, are actually not realistic for masses around 1.5 $M_{\odot}$, as they imply unusual conditions for the carbon isotope ratio and appear in conflict with the  fluorine data. We therefore suggest that the common idea of a noticeable decrease of Li along the AGB is not a viable explanation of the data. Models with moderate mixing rates, evolving at almost constant Li (before production at TDU starts) should be preferred. The spread of the abundances left behind by RGB phases is essentially maintained (with only a marginal reduction) through the early-AGB stages. The real trend was observationally masked by the selection effects introduced by TiO bands, which made so difficult to measure low Li abundances in cool O-rich giants and hence generated the idea that Li-rich K stars could evolve to Li-poor TP-AGB stars (especially if C-rich). Actually, this reversal of the common ideas, which remains tentative (due to the low statistics) for O-rich AGB stars, will be soon shown to be somewhat more robust for C(N) stars.

\section{Modeling Li abundances in C-rich AGB stars}

\begin{table}[h!!]
\centerline{Table 2}
\centerline{Initial conditions adopted on the early-AGB in Figure 10}
\centerline{(2 M$_{\odot}$ models of solar metallicity)}
\centerline{
\begin{tabular}{l|c|c|c|c|c}
\hline
Line & $\Delta_{AGB}$ & $\dot M_{6,AGB}$ & \multicolumn{3}{c}{$^{12}$C/$^{13}$C
after the RGB} \\
\hline
 &  &  & d & c and d & e and f \\
\hline
 continuous & 0.22 & 1 & 8.71 & 8.70 & 8.64 \\
 dashed & 0.22 & 0.3 & 27.10 & 23.78 & 22.49 \\
 short-dashed & 0.15 & 0.1 & 158.80 & 60.15 & 49.42 \\
 dotted  & 0.15 & 0.3 & 120.81 & 72.23 & 62.15 \\
 dash-dotted & 0.1 & 1 & 127.04 & 116.16 & 111.80 \\
\hline
\hline
Case & X($^3$He) & X($^7$Li) & $\Delta$ ; $\dot M_{6,RGB}$ & \multicolumn{2}{c}{FDU
Estimate} \\
\hline
d & 2.09$\cdot 10^{-4}$ & 1.80$\cdot 10^{-10}$ & -- ; -- & \multicolumn{2}{c}{FRANEC
2.5$M_{\odot}$}\\
e & 9.50$\cdot 10^{-5}$ & 3.44$\cdot 10^{-11}$ & 0.22 ; 0.1 &
\multicolumn{2}{c}{FRANEC} \\
f & 9.83$\cdot 10^{-5}$ & 4.07$\cdot 10^{-12}$ & 0.22 ; 0.1 &
\multicolumn{2}{c}{\citet{chala10}} \\
g & 3.35$\cdot 10^{-5}$ & 1.64$\cdot 10^{-12}$ & 0.22 ; 0.3 &
\multicolumn{2}{c}{FRANEC} \\
h & 3.40$\cdot 10^{-5}$ & 1.19$\cdot 10^{-13}$ & 0.22 ; 0.3 &
\multicolumn{2}{c}{\citet{chala10}} \\
\hline
\end{tabular}
}
\label{ini1p5}
\end{table}

More stringent constraints can be derived from more massive stars, which cross both the O-rich and the C-rich phases. As a prototype
for them we use our model with $M = 2 M_{\odot}$. Here an important issue is the moment in which the envelope becomes C-rich due
to TDU. As this moment is known to depend strongly on metallicity (C-rich phases being favored at lower [Fe/H] values), and
as the observed C stars of our sample cover a range in [Fe/H] (down to about $-$0.4) we must consider models allowing for at
least a moderate spread in composition. We therefore refer to stellar evolutionary calculations for solar and half-solar metallicity.

Again, the initial conditions of our AGB calculations have to reproduce the spread observed and modeled at the end of the RGB. The comparison of models and observations is shown in Figure 10 for the two metallicities chosen. Since in our RGB calculations for a 2$M_{\odot}$ star we found a larger range of possible final abundance combinations than in the 1.5$M_{\odot}$ case, we show in Figure 10 more model curves than in Figure 7. Table 2  illustrates, in cases from $e$ to $h$, the values of the Li and $^3$He abundances left by the occurrence of extra-mixing on the RGB, as computed with the parameters indicated in the lower panel of the Table. The effects of extra-mixing are known to decrease for increasing stellar mass (see e.g. Paper 1); we therefore also include a test case (case $d$) in which extra-mixing on the RGB was excluded. On the basis of what we said in Section 2 about the  masses that homogenize the radiative region only after the end of the RGB phases, this test model is actually a simulation of what should occur for $M >$ 2.3$M_{\odot}$. In order to derive the initial conditions on the AGB for this test, we used the previous evolution of a case with $M$ = 2.5$M_{\odot}$ as representative of the masses in this range. All the cases shown have then extra-mixing computed on the AGB for circulation rates from small to large ($\dot M_6$ = 0.1, 0.3, 1). When case $d$ is run with the smallest value of $\dot M_{6,AGB}$, it roughly corresponds to a star that is not affected by deep mixing throughout its evolution. For all the curves, the blue parts correspond to an O-rich envelope, the red ones to C-rich conditions.

\begin{figure}[h!!]
\centerline{{\includegraphics[height=7.8cm, width=7.5cm]{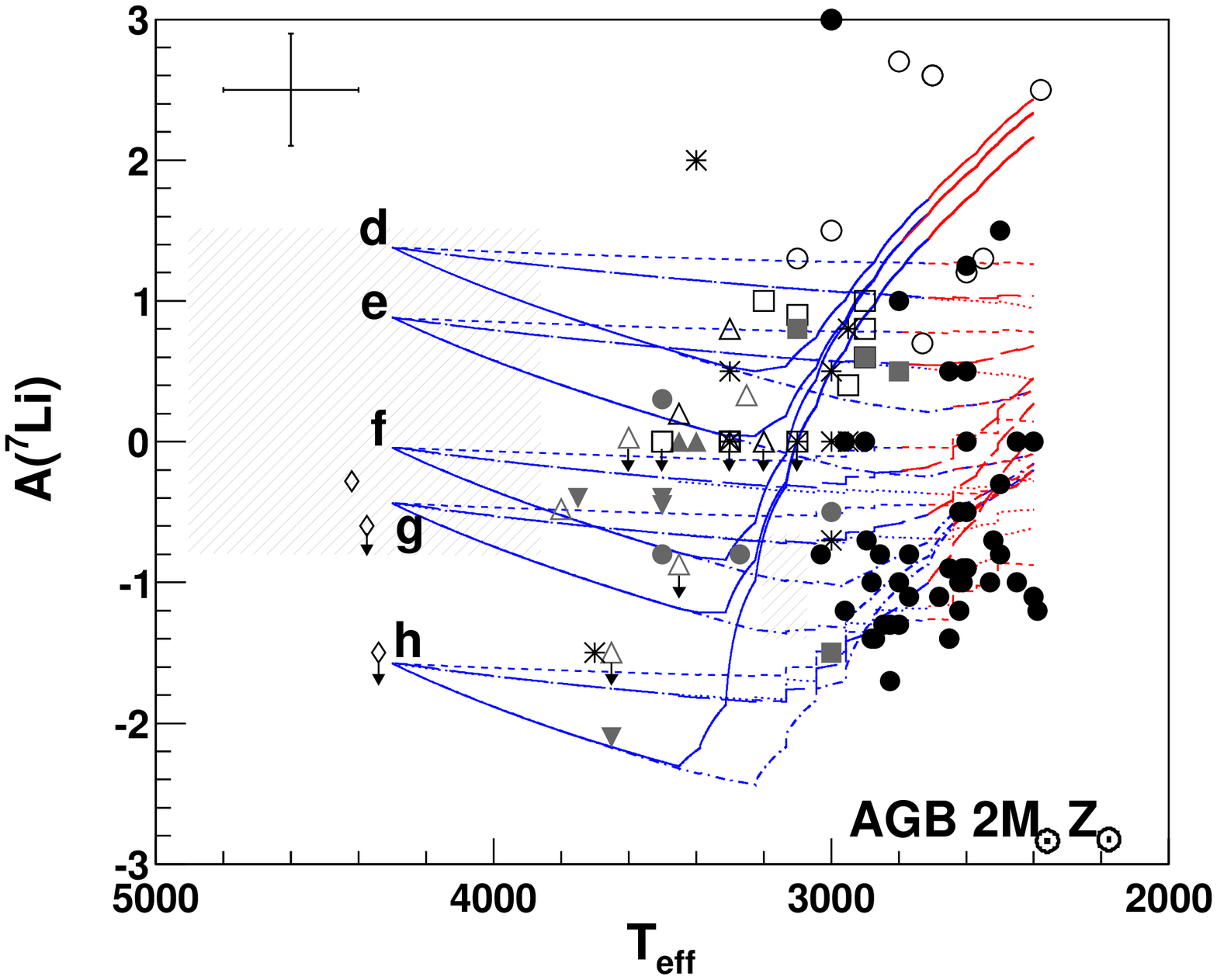}}
{\includegraphics[height=7.8cm, width=7.5cm]{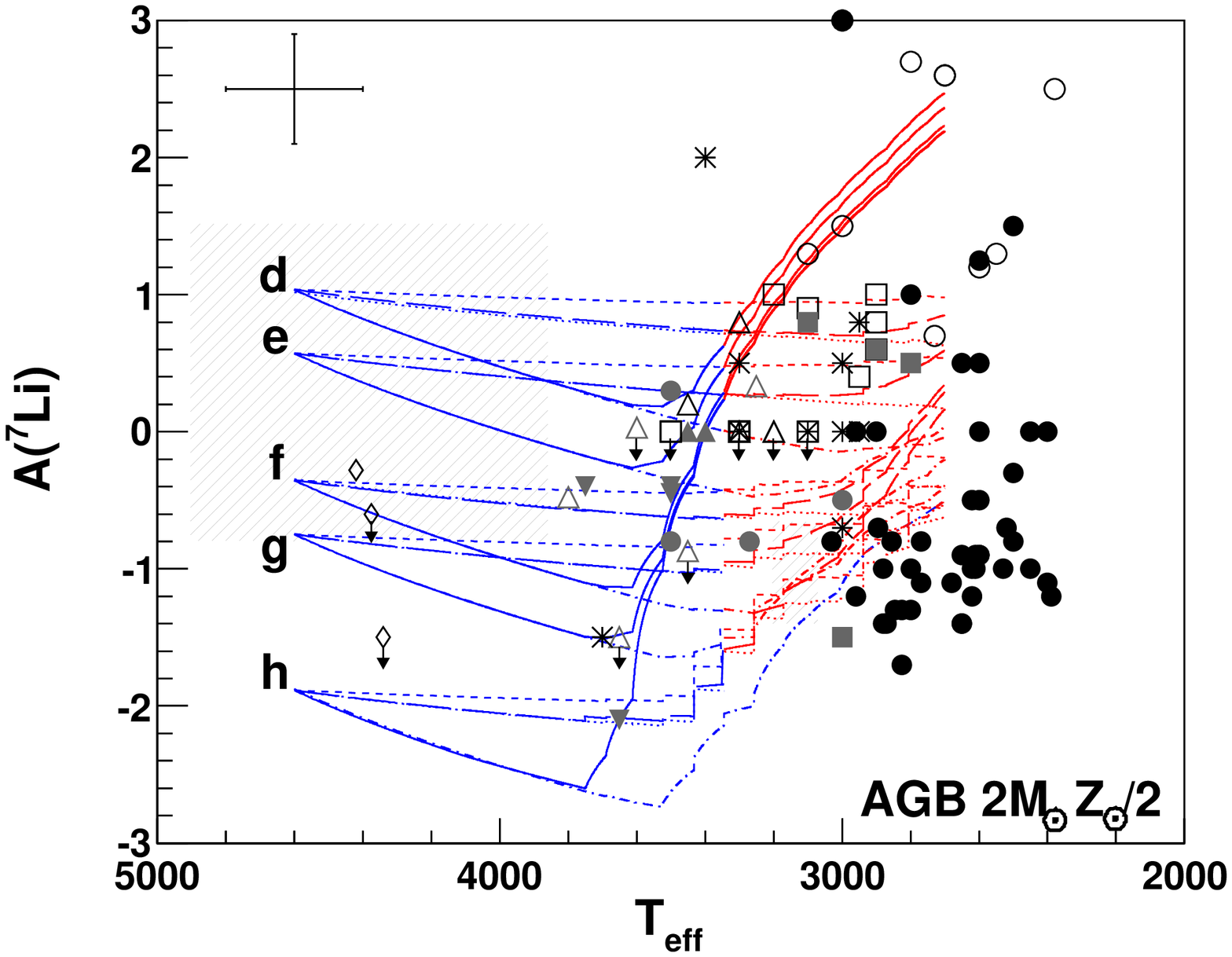}}}
\caption{The left panel shows a plot similar to Figure 7, but including data points for C-rich stars, indicated as circular dots. Open, grey and black dots refer to CJ, SC and C(N) stars, respectively \citep[see][]{denn,abia}. The curves refer to our 2$M_{\odot}$ star of solar metallicity: the blue parts have an O-rich  envelope, the red parts a C-rich one. Right panel: same as the left one, but with models from our 2$M_{\odot}$ star of half-solar metallicity. Note how the C-rich phases extend to rather early TDU episodes, and how the effective temperature extends to higher values at lower metallicity.}
\label{lifig10}%
\end{figure}

Figure 10 shows that the agreement between models and observations is good: cooler C-stars are better
fitted by the solar metallicity model; warmer ones by the case with a lower metal content. Notice that all C(N) giants, at least in the right panel, occupy the region covered by red curves, referring to C-rich conditions. The spread in metallicity chosen should
therefore be representative of the real conditions for these sources).

Despite the general agreement between models and observations, confirming what was already shown for the 1.5$M_{\odot}$ case, the AGB phases do not add much to the spread of abundances present after the RGB. At moderate transport rates the Li consumption on the early-AGB is again small, so that we cannot establish an evolutionary link between Li-rich K or M giants and Li-poor C(N) stars. On the basis of the discussion of Section 4, where we illustrated how the problem of Li re-production from relatively fast mixing is especially severe in 2$M_{\odot}$ models, we can now confidently conclude that such an evolutionary link is physically impossible. Most AGB stars evolve at roughly constant Li from K to late-S types, and even to early-C spectral types. Subsequent production by TDU occurs with an efficiency that depends on the $^3$He and Li abundances in the envelope at the TP-AGB stage, which in their turn are a consequence of the history of mixing in the previous evolution of the star. Hence Li-poor C(N) stars descend from Li-poor M giants; we don't see many such objects because of the selection effects induced by contaminations from TiO bands, as already discussed.

The fact that Li-rich K and M giants of the early-AGB phase are much more numerous than Li-rich C(N) stars seems actually to contradict this evolutionary path at almost constant Li. However, we remember that a way for maintaining a high Li content in early-AGB phases is to suppress extra-mixing on the RGB. We can therefore speculate that Li-rich M, MS and S stars are actually of relatively high mass (e.g. above 2.3$M_{\odot}$, see Section 2). Only the less massive among them might become C rich, because TDU might not be efficient enough to bring the C/O ratio to values larger than 1 in too massive envelopes (say, for $M > 3M_{\odot}$ stars).
We have to admit that this is for the moment a guess: computing models for these masses involves large uncertainties, as we do not have enough observational data to firmly constrain the efficiency of TDU in AGB stars of intermediate mass. Anyway, if our hypothesis is true, the number of Li rich C(N) stars would be reduced because many rather massive, Li-rich objects would not achieve the C-rich stage. There would be therefore only a specific mass interval to form C(N) stars, with a lower limit near 1.7$M_{\odot}$, as shown in Paper 1, and an upper limit possibly around 3 $M_{\odot}$, as required by the low number of them that are Li-rich. Some C-rich giants with high Li and high $^{13}$C, especially of CJ type, can however also derive from another scenario. As the figure shows, several CJ stars  are fitted by fast mixing models, which produce Li as discussed in Section 4. They would not be objects that preserved their original FDU content of Li; on the contrary, they would have destroyed it and subsequently reproduced it, as shown in efficiently mixing models, starting with the low Li abundances of cases $g~$ and~ $h$.

\begin{figure}[h!!]
\centerline{{\includegraphics[height=0.7\textheight]{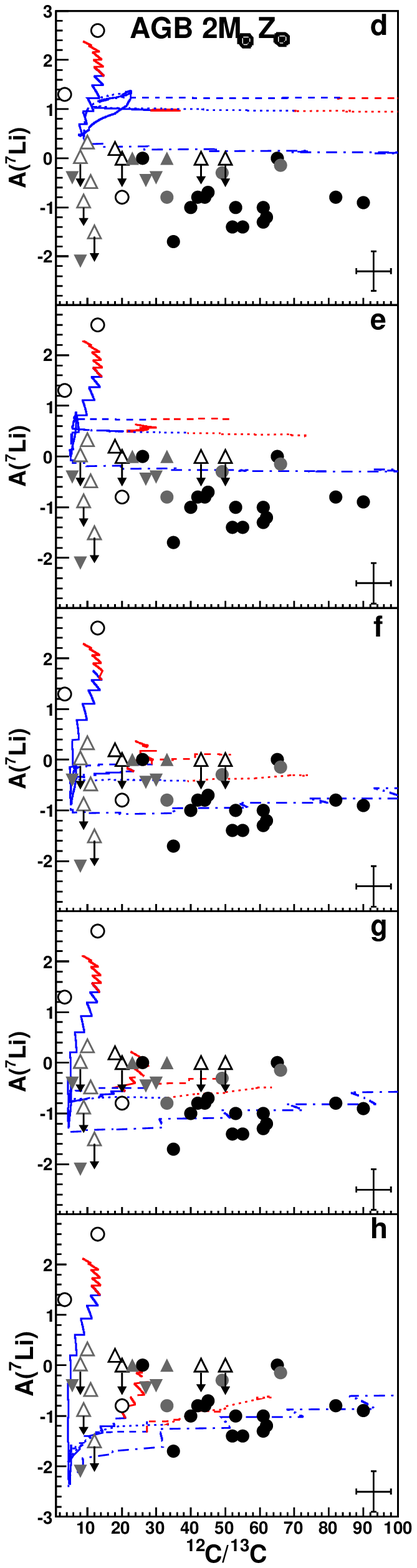}}
{\includegraphics[height=0.7\textheight]{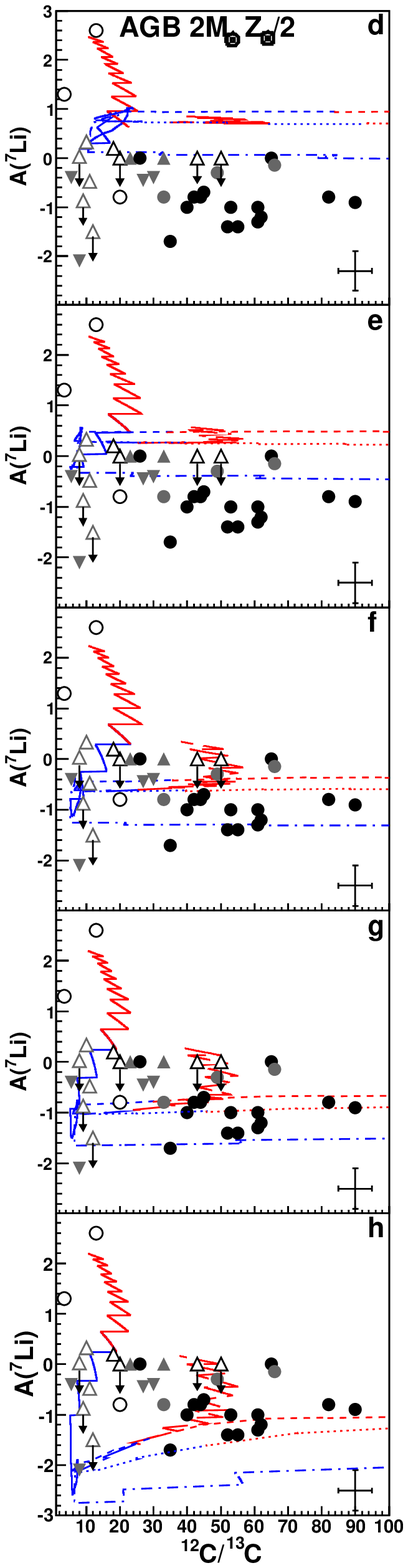}}}
    \caption{Li abundances in O-rich and C-rich AGB stars as a function of the $^{12}$C/$^{13}$C ratio. The symbols and the curve models have the
    same meaning as in Figure 10. Observations of carbon isotopic ratio are from \citet{JSL92} for O-rich giants and from \citet{abia2002} for C-rich stars. The
    left and right panels present models for solar and half-solar metallicity, respectively, starting from the $d, e, f, g$ and $h$ positions reported in Table 2.}
\label{lifig11}%
    \end{figure}

\begin{figure}[h!!]
\centerline{{\includegraphics[height=0.7\textheight]{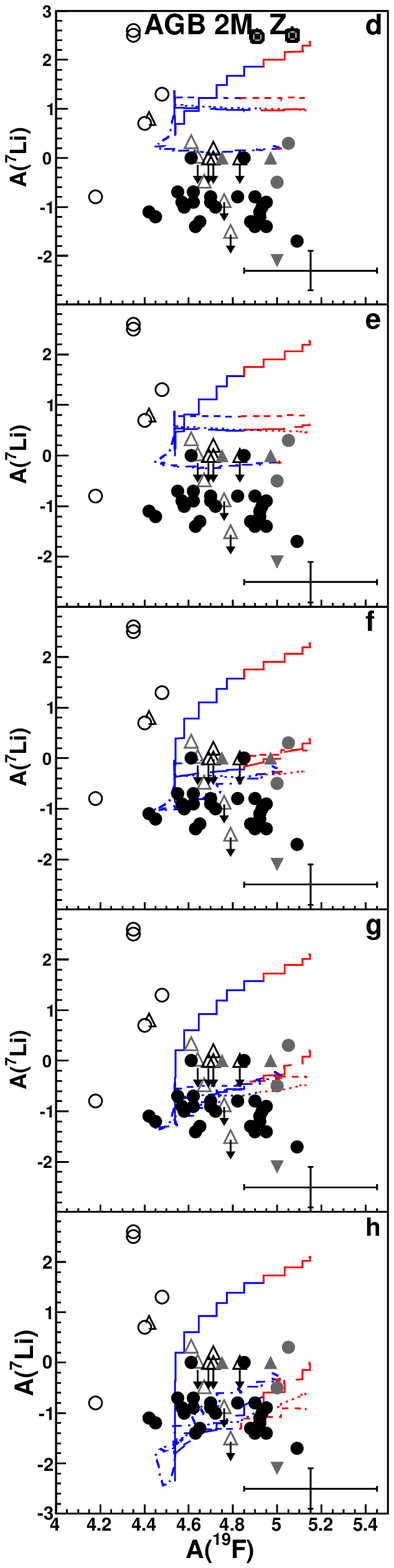}}
{\includegraphics[height=0.7\textheight]{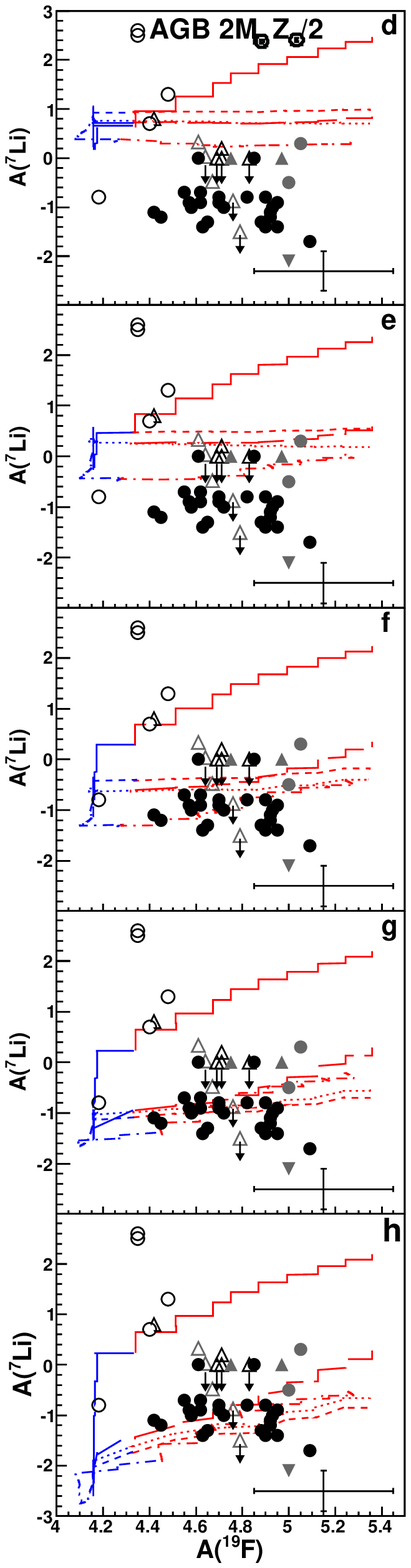}}}
    \caption{Li versus F abundances in O-rich and C-rich AGB stars. The symbols and the
    curve models have the same meaning as in Figure 10. Observations of F ratio are from \citet{JSL92} and \citet{Utt08} for O-rich giants and from \citet{abia2} for C-rich stars. Again, the
    left and right panels present models for solar and half-solar metallicity, respectively, starting from the $d, e, f, g$ and $h$ positions reported in Table 2.}
              \label{lifig12}%
    \end{figure}

Figure 11 illustrates the relations between Li abundances and the $^{12}$C/$^{13}$C ratio. It is evident that the carbon isotope ratio is dispersed over a wide interval, with values ranging from less than 10 to more than 90. As already mentioned, there is not a real correlation, but rather a reduction of the spread in Li abundances for increasing values of the $^{12}$C/$^{13}$C ratio. Notice again that CJ stars (open circles) are not well reproduced by the normal models with slow mixing: a few of them lay close to some theoretical tracks, but generally only for the O-rich parts of them (blue curves). They instead fit well with the cases with a high transport rate; this is so, inside the errors, also for a couple of very Li-rich objects of this class at the upper left. Both their low carbon isotope ratio and their high Li abundance are actually permitted by a fast mixing ($\dot M_6 \gtrsim$ 1), as already discussed. Indeed, this process easily consumes $^{12}$C, producing $^{13}$C; it also produces Li (see Section 4). Hence, as speculated for Figure 10, CJ stars must be the hosts of fast mixing processes.
The remaining C stars, mainly of C(N) type, are instead well accounted for with moderate circulation rates ($\dot M_6 \lesssim 0.3$). In these cases, models with a lower metallicity compare better with stars with very low $^{12}$C/$^{13}$C values.

Summing up, the data are reproduced rather well, but their spread corresponds to a wide range of possible situations, making predictions uncertain. It is in any case significant that models $g$ and $h$ (with little Li at the beginning of the AGB) are those that fit the largest part of Li-poor C(N) stars; they also account for the observations of Li-rich CJ stars, but in this case require fast mixing on the AGB. In general, C stars need the occurrence of deep mixing episodes throughout their whole evolution (even on the MS, if case $h$ has to be trusted).

From what has been discussed so far, it is not surprising that Li abundances in AGB stars have been so far difficult to understand: each star owes its Li content to the integral of the effects accumulated in its previous evolution, so that minimal differences in the mass, or rotation rate, or evolutionary status, or mass loss, etc., combine in yielding a zoo of different and puzzling situations. Figures 10 and 11 tell us that these situations can actually be understood; but unfortunately they also require that different stars experience mixing with a range of different values for the controlling parameters. While this limit still allows the models to interpret (a posteriori) the data, it prevents safe predictions on the properties of the parent stars.

Finally, Figure \ref{lifig12} shows the comparison between Li and F abundances for O-rich and C-rich stars together, superimposed to our predictions from nucleosynthesis calculations in 2$M_{\odot}$ stars. It is clear that our Population I models, accounting for a small spread in metallicity and (in most cases) moderate rates of
non-convective transport, provide a good understanding of the observations. In the complex scenario we have outlined so far, this agreement is important for us, as the
data refer to elements produced independently: one (Li) above the H- burning shell, the other (F) in the He- and C-rich layers, during TPs. Again, most observations are
bracketed by models of types $g$ and $h$, implying extra-mixing also in previous evolutionary stages, possibly down to the MS. In general, in Figure 12 the cases
with moderate values of the $\dot M_6$ and $\Delta$ parameters leave the F abundance dredged up from the He-rich layers essentially untouched, which fact seems actually
to be required by the observations. On the contrary, the models characterized by fast mixing do not fit well C(N) stars. If they reach hot temperatures (dash-dotted
lines) they undergo F and C destruction during the O-rich phase. At solar metallicity, they barely reach the C/O=1 condition near the AGB end, failing to
reproduce the C-rich phases of Figure 12 (left panel). For a lower content of heavy elements (right panel) they can reach C/O $>$ 1 phases, but in so doing they
miss more than 50\% of the data points for F. If instead fast mixing cases reach only moderate temperatures (continuous lines), then they proceed at too high Li
abundances (and indeed reach up to the high Li concentrations of some CJ stars, as already mentioned). Notice that in so doing they also pass through the area where
some F-poor stars of CJ type lay. Hence, fluorine seems to offer some independent (although not very strong) indications that efficient mixing phenomena, while
incompatible with C(N) observations, might instead be more common in CJ stars.

\section{Discussion and Conclusions}
In this paper we have analyzed those evolved red giants that generally show Li abundances $A$(Li) $\lesssim$ 1.5 in their spectra, i.e. envelope concentrations lower than (or equal to) the typical maximum values provided by stellar models after the FDU. We attributed the Li consumption to the occurrence of extra-mixing mechanisms, which we parameterized, in line with what was done in Paper 1, on the basis of the efficiency of mass transport and of the maximum temperature of the mixed layers. At solar metallicity, circulation rates from low to moderate ($\dot M_6 = 0.03 - 0.3$) are sufficient to account for the observations of Li and $^{12}$C/$^{13}$C ratios in RGB stars. This would roughly correspond, in a diffusive treatment, to diffusion coefficients $D_{mix}$ of the order of 3$\times10^6$ to 3$\times$10$^7$ cm$^2$/sec, i.e. on average a factor-of-ten smaller than found by \citet{dw} for Population II stars. This might be a confirmation of the known decrease of deep mixing efficiency with increasing metallicity.

The reproduction of the Li observations in red giants requires consideration of a rather large spread of Li abundances at FDU, as can be found in the current literature. These abundances, in their turn, imply that non-convective mixing has previously operated on the MS at different efficiencies, probably due to rotational effects, thus dispersing the Li abundance that is induced in the envelope by FDU, which would be otherwise confined in a small range ($A$(Li) = 1 $-$ 1.5).

The occurrence of extra-mixing in RGB phases is possible after the advancement of the H-burning shell has erased the
chemical discontinuity created by FDU. With the reaction rates revisions discussed in Paper 1 we find that only in stars below about 2.3$M_{\odot}$ this can be obtained before the RGB tip. More massive stars, therefore, cannot experience deep-mixing episodes during their ascent to the RGB. On the other hand, during core He-burning and through the major part of the ascent to the early-AGB, the envelope remains rather far from the H-burning shell; any extra-mixing phenomenon is therefore bound to restart only later, when the star is again close to or directly on the Hayashi track. This includes the last $\simeq$ 1 Myr of the early-AGB and the subsequent TP-AGB. Globally, the  available time (about 4 Myr) is typically a factor-of-ten shorter than for the RGB, so that if extra-mixing occurs at the same small or moderate rates, little Li is destroyed. This fact is in agreement with what was shown by \citet{guan09}, although in their analysis the subsequent Li production from TDU is prevented by the very low abundance of $^3$He they considered. Conversely, when convective TDU episodes start, they reproduce Li, rapidly saving to the envelope whatever $^7$Be has remained above the H shell. On the other hand, we showed that any faster extra-mixing episode (at $\dot M_6 \gtrsim$ 1) does not destroy more Li, but actually re-produces it, because its overturn time scale becomes, at least in the external layers of the radiative zone, faster than that for Be decay.

We therefore conclude that effective Li destruction on the AGB, suitable to establish an evolutionary link between Li-rich K or M giants and Li-poor C(N) stars is not physically possible. The common interpretation that a considerable Li destruction must occur in between these evolutionary stages is wrong. On the contrary, stars evolve along the AGB, before the TPs, at almost constant Li. Li-poor C(N) giants must therefore descend from Li-poor early-AGB stars. These last are observed in very limited numbers only because of the presence of TiO bands, preventing the measurements of low Li abundances, at least in cool M giants. Conversely, Li-rich M giants (statistically much more abundant that Li-rich C(N) stars) should owe their high Li content to the fact that their mass is high enough to prevent extra-mixing to occur on the RGB ($M > 2.3 M_{\odot}$). Only the less massive among them will become C-rich, the others will remain O-rich (due to their too large envelope masses, for which TDU cannot induce the C/O $>$ 1 condition). Hence, two different phenomena conjure in creating strong selection effects in the comparisons between O-rich and C-rich AGB stars, inducing the false appearance of a reduction of Li along the AGB.

The above scenario is confirmed by the limited data available for Tc (in the sense that, with the moderate extra-mixing efficiencies mentioned, all Tc-rich AGB stars are fitted by models that are crossing the TP-AGB stage, where Tc is mixed to the envelope by TDU).

A more quantitative check is possible thanks to F abundances. Extra-mixing at rapid rates and reaching down to hot temperatures destroys F, while most C(N) stars appear to be F-rich. Hence, the trend of fluorine confirms that only moderate transport rates (leaving unchanged the abundances of F dredged up by TDU) must characterize the majority of C(N) stars. In this scenario, CJ stars seem to be those few objects that experienced fast mixing: its occurrence indeed produces Li, and they are Li-rich; it destroys $^{12}$C producing $^{13}$C, and they have low $^{12}$C/$^{13}$C ratios. It also destroys F, and CJ stars are known to be F-poor.

\begin{figure}[h!!]
\centerline{{\includegraphics[height=7.8cm, width=7.5cm]{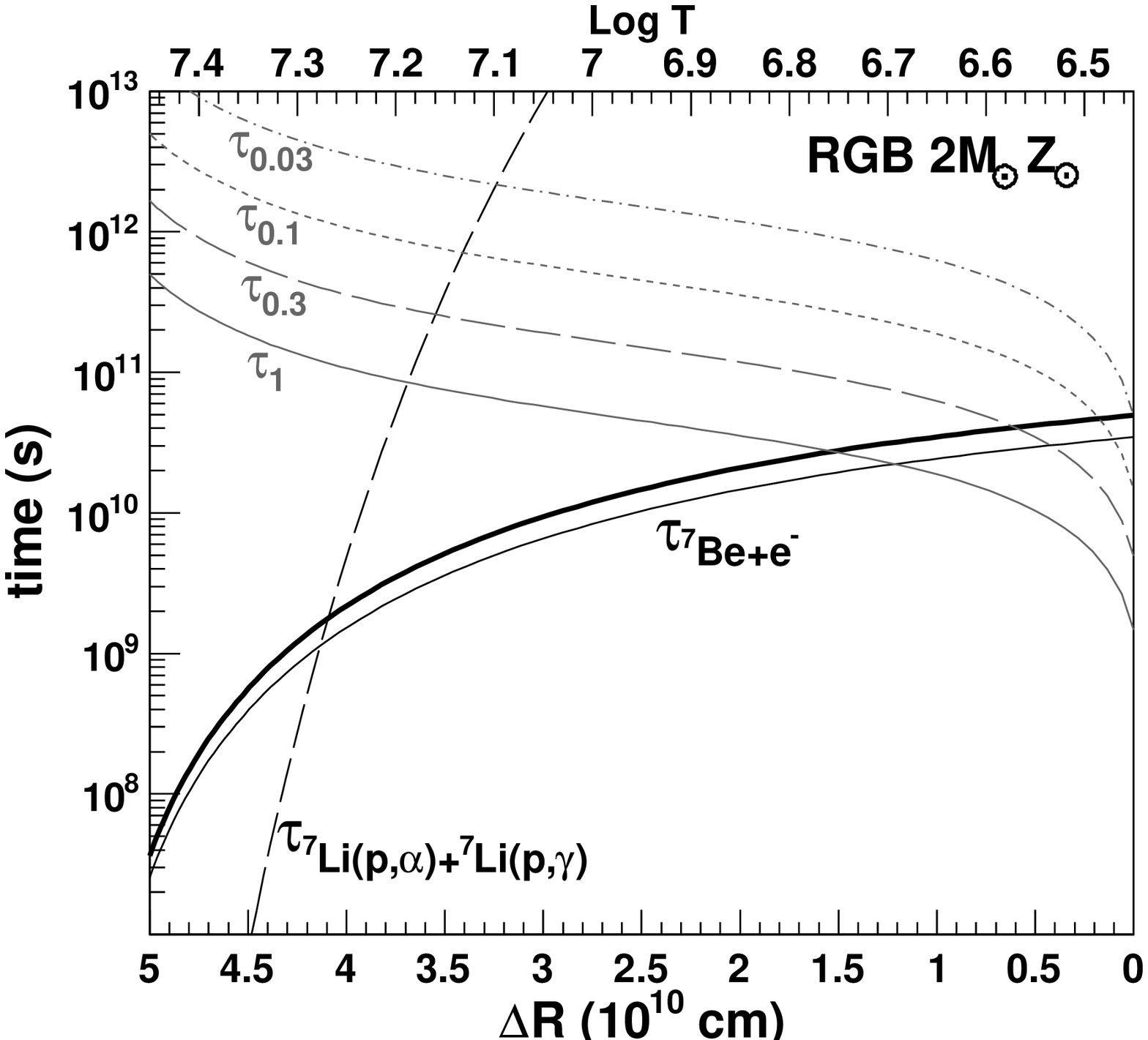}}
{\includegraphics[height=7.8cm, width=7.5cm]{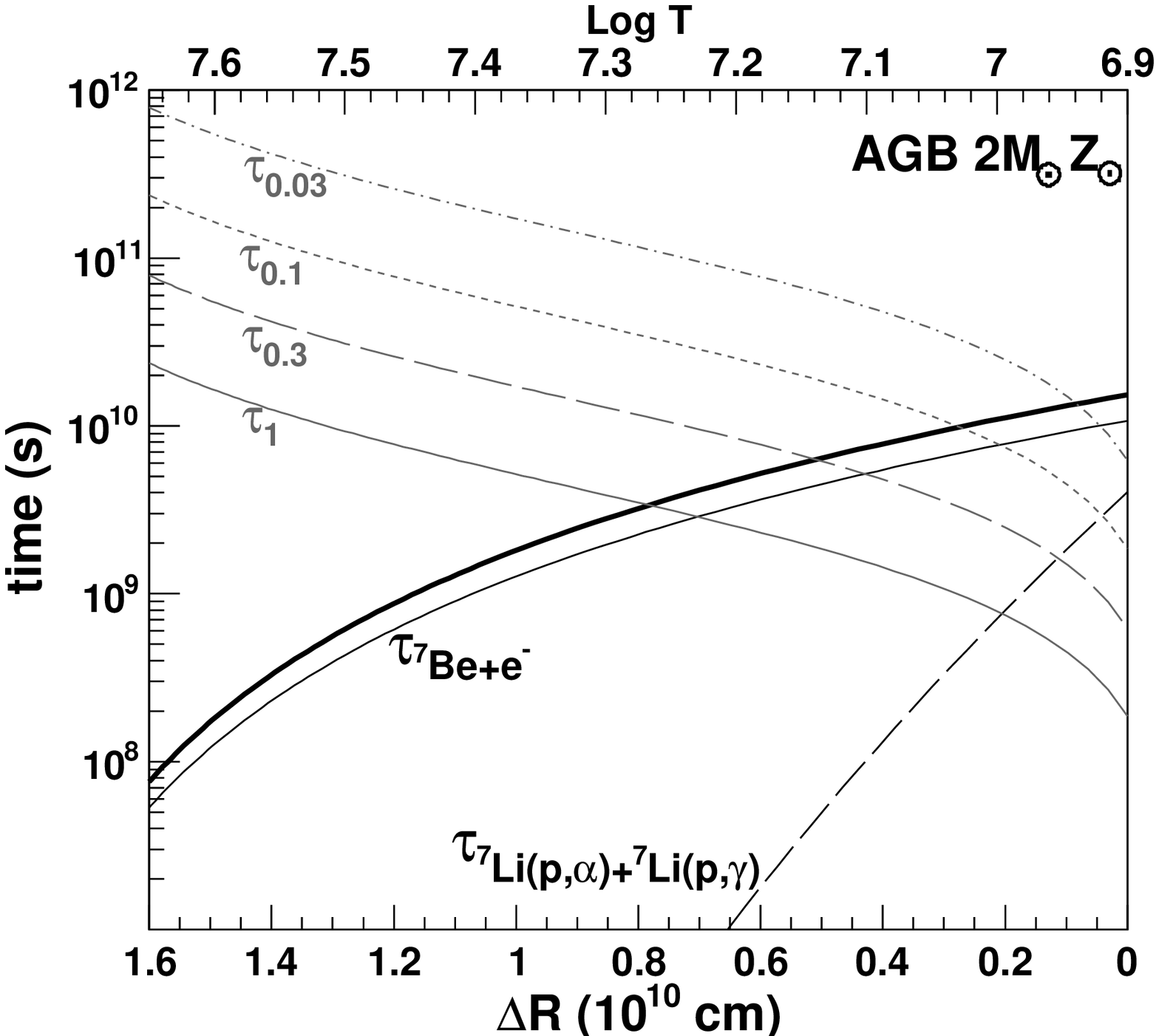}}}
    \caption{The effects of a reduction by 30\% of the rate for Be decay (bold continuous curve),
    as compared to the presently accepted value (normal continuous curve). The allowed variation
    accounts for the uncertainties mentioned for this rate in Appendix I. As is clear from the
    plot, the balance between Li destruction and production, due to the competition of the time scale
    for mixing with that for decay, is changed only partially on the AGB (right panel) and even less on the RGB (left panel).}
              \label{lifig13}
    \end{figure}

All the above conclusions are obviously tentative, as the interpretation of the data is not straightforward and the statistics is not large. Cautions are also due to the uncertainties affecting
the rate for $^7$Be decay, as discussed in the Appendix. One has to notice that the H-burning shell in evolved stages has a higher temperature and a lower density than the central regions of the present Sun.
Hence the plasma conditions should make Be decay through the capture of free electrons even more difficult than in the Sun. Obviously, a longer Be lifetime would require changes in our discussion about the competition between the time scale for decay and that for mixing overturn. However, if very special effects in the plasma (see below) can be neglected, then the effects of a possibly longer half-life for $^7$Be are not expected to be dramatic. In order to give a more quantitative support to this statement, Figure 13 (bold continuous line) shows the effects of an increase by 30\% of this half life, according to the suggestions by \citet{shsh}; the presently accepted value is shown by a normal continuous line. Figure 13 refers to the most critical case, that of a 2$M_{\odot}$ star. It shows that the change does  modify drastically the competition with the overturn time scale for mixing on the RGB; on the AGB it is more relevant, implying an increase of the effective area for Li production (where $\tau_{mix} \lesssim \tau_{e^-}$), which is however not particularly crucial, given all the other uncertainties present. The main source of difficulty in modeling the Li abundances in red giants seems therefore to be linked to the astrophysical scenario, not to the nuclear input data. A different situation might occur if the presence of magnetic fields can modify the energy distribution of the particles inside the plasma, inducing Lorentz forces.  However, no nuclear physics calculation seems to have considered this possibility so far, so that the rate for Be decay in such a situation remains essentially unknown. A more complete analysis of this issue will be performed in the future paper dedicated to Li-rich stars.

\newpage

{\bf APPENDIX I Nuclear Processes Controlling the Nucleosynthesis of Li}

 Most nuclear reaction rates used in the present work are based on a recently published review of solar fusion cross sections \citep{ade11}. For some reactions not included there, the recommendations by NACRE \citep{angulo} are adopted.

We report here a brief analysis of the present situation, underlining  the uncertainties affecting nuclear data that might have an impact on the synthesis of lithium in stars.

{\bf 1. $^2$H(p,$\gamma$)$^3$He and $^3$He($^3$He,2p)$^4$He.}
The recommendations by \citet{ade11} for the astrophysical S-factors of these reactions are S$_{12}$(0) = 0.214$^{+0.017}_{-0.016}$ eV b and S$_{33}$(0) = 5.21$\pm$0.27 MeV b, respectively.
These reactions have been well constrained experimentally down to the Gamow peak in the Sun \citep{case,bone} and there appears to be no room for anything different than an almost instantaneous conversion of deuterium into $^3$He and the following production of a rather well defined abundance of $^4$He.

{\bf 2. $^3$He($^4$He,$\gamma$)$^7$Be.}
The recommendation by \citet{ade11} is S34(0)=0.56$\pm$0.02(exp.)$\pm$0.02(theor.) keV b, where the more recent measurements by \citet{singh}, \citet{br07}, \citet{costa}, and \citet{dileva} are considered. A closer look at the data shows a  slight tension between them, possibly indicating some discrepancy due to systematic uncertainties. Recently, a new {\it ab-initio} calculation \citep{neff} has been presented, where a realistic nucleon-nucleon interaction is used to calculate the cross section. The resulting S$_{34}$ is 0.593 keV b. The agreement with the data by \citet{br07} and \citet{dileva} is very good, both regarding the absolute value and the energy dependence of the S factor; instead, the agreement with the results from LUNA \citep{costa} and from the Weizman Institute \citep{singh} is at a 2$\sigma$ level or worse. An extrapolation leading to S$_{34}$=0.59 keV b should therefore be considered as a realistic alternative to the quoted recommended value (this would however imply only an increase by 5\%). In the paper we adopt, for a question of uniformity, the \citet{ade11} choice; the possible increase is too low to have any significant effect.

{\bf 3. $^7$Be(e$^-$ $\nu$)$^7$Li.}
Here the recommendation by \citet{ade11} is essentially based on the
work made by Bahcall and collaborators \citep{bahc69}. In turn, their calculations were based on the paper by \citet{iben}, where a partial ionization of $^7$Be in the Sun was predicted.
Therefore, the rate now adopted includes both a bound state and a continuum contribution to the decay, with a total-to-continuum capture ratio of 1.217. A rather different result was obtained by \citet{shsh}, who predicted a fully ionized condition for $^7$Be in the Sun. The missing contribution from bound electrons to the capture rate was partially compensated by a larger contribution from the continuum. Summing up, an increase of the lifetime by only 20 to 30\% resulted, as compared to the above-mentioned recommendations. The situation is now even more complicated, as \citet{qs9} recently found a $^7$Be lifetime shorter by about 10\%, due to the use of a modified Debye-Hueckel screening potential. On the contrary, \citet{bahc} excluded the validity of a modified Debye-Hueckel potential for a dense stellar plasma. In conclusion, the real questions in this case are: i) which is the degree of ionization of $^7$Be in a stellar plasma, e.g. in the Sun core? ii) Further, as the $T$ conditions above the H-burning shell in evolved stars span a large range (from a few $\times$10$^7$ to a few $\times$10$^6$ K) and as the density is lower than in the Sun, how does the $e^-$-capture rate react? (H-shell conditions, in particular, might modify considerably the equilibrium between bound and free electrons, and even the
number of free electrons contributing to the capture). The large uncertainties, both of observational and theoretical nature, affecting Li abundances in stars have also to include this poor understanding of the basic nuclear input data.

{\bf 4. $^7$Be(p,$\gamma$)$^8$B.}
In this case the recommended value by \citet{ade11} is $S_{17}$(0) = 20.8$\pm$7(exp.)$\pm$1.4(theor.) eV b.
This value is estimated considering only direct measurements (whose results show some mutual discrepancy); indirect (Coulomb dissociation) measurements were not taken into account. The comparison between direct and indirect measurements is debated \citep{esbe,gai06}, but the overall impact on the $^7$Li abundance should be negligible, when compared to the uncertainty deriving from the reactions $^3$He+$^4$He and $^7$Be+e$^-$.

{\bf 5. $^7$Li(p,$\gamma$)$^8$Be.}
This reaction has not been included in the compilation by \citet{ade11}, so we use the NACRE recommendation \citep{angulo}, which is $S_{71\gamma}$(0)= 0.3 keV b. The measurements by \citet{spraker} later suggested a doubling of this rate; however its impact on Li destruction is minimal, due to the prevailing role of the (p,$\alpha$) channel.

{\bf 6.  $^7$Li(p,$\alpha$)$^4$He.}
Also for this reaction, we use the NACRE value ($S_{71\alpha}$(0)= 55.6$^{+0.8}_{-1.7}$ keV b, later confirmed by \citet{cruz}.
One should note, however, that an anomalous screening is observed in this case, whose origin is not well understood. This might have an impact on the stellar rate.

{\bf Acknowledgements} We are indebted to S. Randich and L. Magrini for many useful discussions on spectroscopic data for Li abundances in open
clusters and in the Sun. We then owe a lot to OUR longstanding collaborations with R. Gallino, K.M. Nollet, L. Piersanti, O. Straniero and G.J. Wasserburg. S.P and M.B. are grateful to Group III of INFN for support. C.A. and S.C. were partially supported by the Spanish grants AYA2008-04211-C02-02 and FPA2008-03908 from the MEC. S.U. acknowledges support from the Austrian Science Fund (FWF) under project number P~22911-N16.

\end{document}